\documentclass{article}
\usepackage{url}
\usepackage{cite}
\usepackage{chapterbib}
\usepackage{amsmath,bm}
\usepackage{amssymb}
\usepackage{graphicx,wrapfig}
\usepackage{float}
\usepackage{cite}
\usepackage[hidelinks]{hyperref}
\usepackage[noabbrev,nameinlink]{cleveref}
\usepackage{cutwin}
\usepackage{epstopdf}
\usepackage{array,booktabs}
\newcolumntype{L}{@{} >{\kern\tabcolsep} c <{\kern\tabcolsep}}
\usepackage{colortbl}
\usepackage[dvipsnames,svgnames,x11names,tables]{xcolor}
\usepackage{csquotes}
\reversemarginpar

\usepackage[font=small,labelfont=bf,
   justification=justified,
   format=plain]{caption} 
\begin{document}

\centerline{\Large{\bf Network mechanisms of working memory: the role of neuronal nonlinearities}}\mbox{}\\

\centerline{\large{Alex Suarez-Perez$^*$, Omri Harish$^*$, David Hansel}}\mbox{}
\centerline{\normalsize	{$^*$ Joint first authorship }}\mbox{}\\[0.2cm]
\centerline{\large{CNRS UMR 8002 and Universit\'{e} Paris Descartes }}\mbox{}\\
\centerline{\large{45 rue des Saints P\`{e}res, 75270 Paris, France}}\mbox{}\\[0.2cm]

\centerline{{\bf Abstract}}\mbox{}\\
The oculomotor delayed-response (ODR) task is a common experimental paradigm of working memory (WM) study, in which a monkey must fixate its gaze on the center of a screen and, following a brief cue that flashes on the screen, keep fixating for several more seconds before shifting its gaze to the location where the cue flashed. Consequently, in the delay period between the cue and the response the monkey must maintain a memory of cue location. Electrophysiological recordings from the prefrontal area of the cortex (PFC) revealed neurons that display selective persistent activity: their firing rate change induced by the cue persists through delay period, but only in response to a confined range of cue locations. This suggests that the representation of the cue is maintained in the network by a change in network activity profile. In this work, we study a network of rate-model neurons that is capable of preserving information about a past input, owing to structured connectivity and nonlinearities in the neuronal transfer function (TFs). Particularly, we focus on the acceleration of the TF close to firing threshold and the concavity around TF saturation. Any memory mechanism which exploits TF saturation means that some neurons must fire close to their saturation rates; with our model, however, we show that a certain relation between the excitatory and inhibitory neurons' TFs can cause an effective saturation in the network without forcing the neurons into the saturating parts of their TFs. In addition, this mechanism enables the erasure of memory at the end of the delay by a global excitatory signal. Finally, we demonstrate the mechanism in a model network of spiking neurons which describes with more detail the oscillatory dynamics in the state transition due to the interaction of membrane and synaptic time constants which is neglected in the rate model.
\\

{\bf Keywords:} Visuospatial working memory, persistent activity, direction selectivity, computational model, neuronal transfer functions, recurrent connectivity
\clearpage
\section*{Introduction} \label{W1_Intro}

Working memory (WM), the ability to temporarily hold, integrate, and process information to produce goal-directed behavior, is crucial to higher cognitive functions such as planning, reasoning, decision-making, and language comprehension \cite{Baddeley1986,Fuster2015}. The persistent activity recorded in neocortex during WM tasks is thought to be the main neuronal correlate of WM \cite{Fuster1971,Miyashita1988,Goldman-Rakic1995}. For example, in an oculomotor-delayed response (ODR) task in which a monkey has to remember the location of a stimulus for several seconds to make a saccade in its direction, a significant fraction of the neurons in the prefrontal cortex (PFC) modify their activity persistently and selectively to the cue direction during the delay period \cite{Funahashi1989,Funahashi1990,Funahashi1991,Constantinidis2001,Takeda2007}. The classical view is that this reflects a multistability in the dynamics of the PFC circuit because sensory inputs are the same in the precue and in the delay periods but neuronal activity is different \cite{Hebb1949,Hopfield1984,Amit1995,Amit1997,Wang2001}.

In monkeys performing an ODR task, neurons in the dorsolateral prefrontal cortex (DLPFC) show elevated activity during the delay period which depends on the direction of the presented cue (e.g. \cite{Constantinidis2001,Constantinidis2004,Funahashi1989,Funahashi1990,Funahashi1991}). As the direction of the cue is varied the activity of the neuron changes. The response of neuron in the DLPFC is therefore characterized by its tuning curve. Neurons are characterized by its preferred direction i.e. the direction for which the response is maximum. In the classical view, the “line of attractors” hypothesis posits that this direction tuning is an emergent property of the recurrent dynamics in the DLPFC. According to this hypothesis the location of the cue is encoded in persistent states, each state characterized by an activity profile in the feature space which is “bumpy”. Since this space has the geometry of a ring, the location of the bump can be parametrized by an angle which match the direction of the cue. Thus, the set of attractors is continuous which is invariant by rotation. Recent experiment results support this “line of persistent attractors” hypothesis (see \cite{Knierim2012,Hulse2020} for a review). The results reported in \cite{Wimmer2014} show that the behavioral error patterns in ODR tasks correlate with shifts in the tuning curves of individual DLPFC neurons on these error trials, implying a drift of an activity bump during the delay period. Persistent activity bumps encoding for a direction were observed in \emph{drosophila} flies using calcium imaging \cite{Seelig2015,Kim2017,Green2017}. 

Non selective persistent activity emerges naturally in unstructured recurrent networks provided the recurrent excitation is strong enough \cite{Wang1999,Hansel2001}. To prevent the activity to blow up in the persistent state a non-linearity is required. Sigmoidal input-output neuronal transfer function provides the network with an appropriate stabilizing mechanism \cite{Brunel2000}. However, in that case, neurons in the persistent state will fire near saturation. Alternatively, the stabilizing non-linearity can result from the recurrent inhibition. This requires the response of the inhibitory neurons to be more sensitive to inputs than the excitatory neurons \cite{Latham2000,Brunel2001}.

On another hand, non-persistent activity selective to a stimulus characterized by a continuous angular variable can emerge from recurrent interactions. This was first investigated in \cite{Ben-Yishai1995}. The “ring model” introduced in this seminal paper and its generalizations provide a classical framework to investigate the role of excitation and inhibition in orientation selectivity (see \cite{Hansel1998,Priebe2016} for a review). Selectivity in the ring model stems from feature specific strong excitatory connectivity stabilized by inhibition. Combined with the non-linearity of the input-output transfer function of the neurons, the network undergoes a Turing instability, as the external input increases, leading to a line of bump attractors \cite{Goldberg2004}. In this mechanism, the bump attractors are not persistent: if the external input which represent the stimulus is withdrawn the bump is abolished.

Compte et al. \cite{Compte2000} showed that persistence and selectivity to direction can emerge in a network with structured connectivity. With this model they studied the involvement of slow excitation via N-methyl-D-aspartate receptor (NMDAR) channels, and showed that a decrease in the percentage of NMDAR channels out of the excitatory synapses can destabilize the representation in the network. 

In this paper we investigate how the interplay between excitation, inhibition and the nonlinearities of input-output neuronal transfer function gives rise to direction selective and persistent delay activity. To this end we consider networks of excitatory and inhibitory neurons with feature specific connectivity. Neurons are modeled as rate units or integrate-and-fire elements. We combine analytical calculation with numerical simulations to characterize the network stable states as a function of the external input, the interactions strength and the spatial modulation. We show that the transient network dynamics during the switch-on and the switch-off of the persistent selective state depends on the nature of the non-linearities in the input-output transfer function of the neurons.

\section*{Results}\label{W1_Results}

\subsection*{A functionally-segregated working memory mechanism}\label{ssect:DivConq}

\begin{figure}[!t]
\begin{center}
\includegraphics[width=6in]{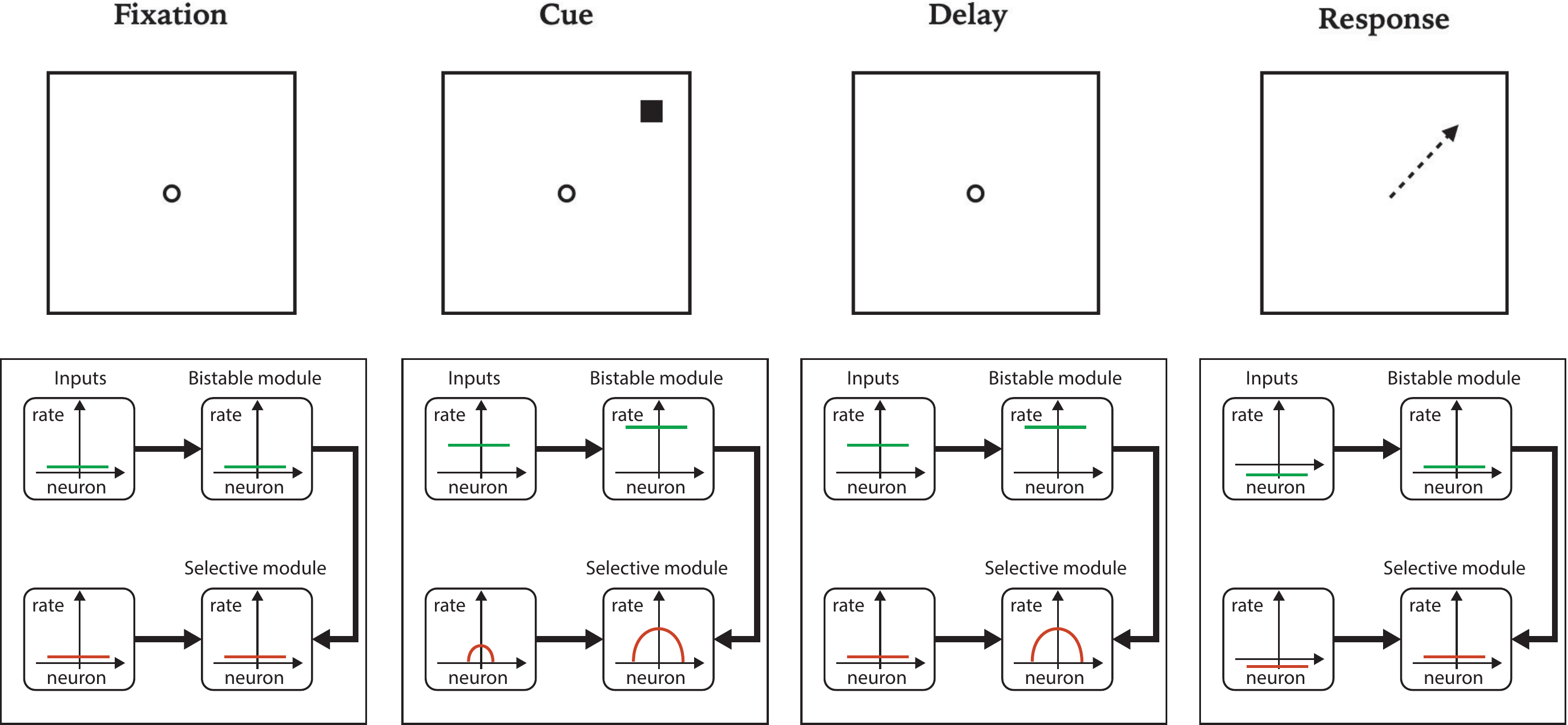}
\end{center}
\caption{\bf Schematic description of the functionally segregated model.} 
\centering
\label{fig:DivConq}
\end{figure}

The first mechanism we describe spatial direction memory involves two modules that are connected in a feedforward manner. The first module is a bistable, non-selective network: there are two possible steady states for this network, and the input to every neuron depends only on the presence of the cue and not on its direction. This module, without storing quantitative information about the cue, functions as a switch between a state in which the memory is empty and a state in which a memory is maintained. The information about the cue direction is in the second module, which is monostable and selective: for every input value there's only one possible steady state, and the input to each neuron is maximal for a specific direction of the cue, termed the neuron's \enquote{preferred} direction (PD). In this module the connectivity is such that if the input is below some threshold the network maintains a low homogeneous state, but if the input is large it settles on a \enquote{bump} state, in which neurons with PDs around a certain direction fire at higher rates than others. The center of this bump is determined either by the center of the stimulus input profile or, if the stimulus input is homogeneous, by asymmetries in the initial conditions. Both modules therefore receive stimulus-related input, but the bistable module receives only information about the existence of a cue whereas the selective module receives information about the direction of the cue. Before the cue period, both modules are inactive (figure \ref{fig:DivConq}, left). During the cue period the bistable module receives a homogeneous stimulus and elevates its firing rate, and the selective module shifts to a bump profile with a center aligned with the stimulus input profile center (figure \ref{fig:DivConq}, middle left). After the cue is removed, the bistable module relaxes to its up-state (rather than back to the down-state),. The selective module, now receiving an above-threshold input, remains in a bump state, thus maintaining a memory of the stimulus input direction during the delay period (figure \ref{fig:DivConq}, middle right). Finally, a negative homogeneous stimulus to both modules brings back the bistable module to the down state and terminates the bump state in the selective module since it is receiving sub-threshold input (figure \ref{fig:DivConq}, right).

\begin{figure}[h!]
\begin{center}
\includegraphics[width=6in]{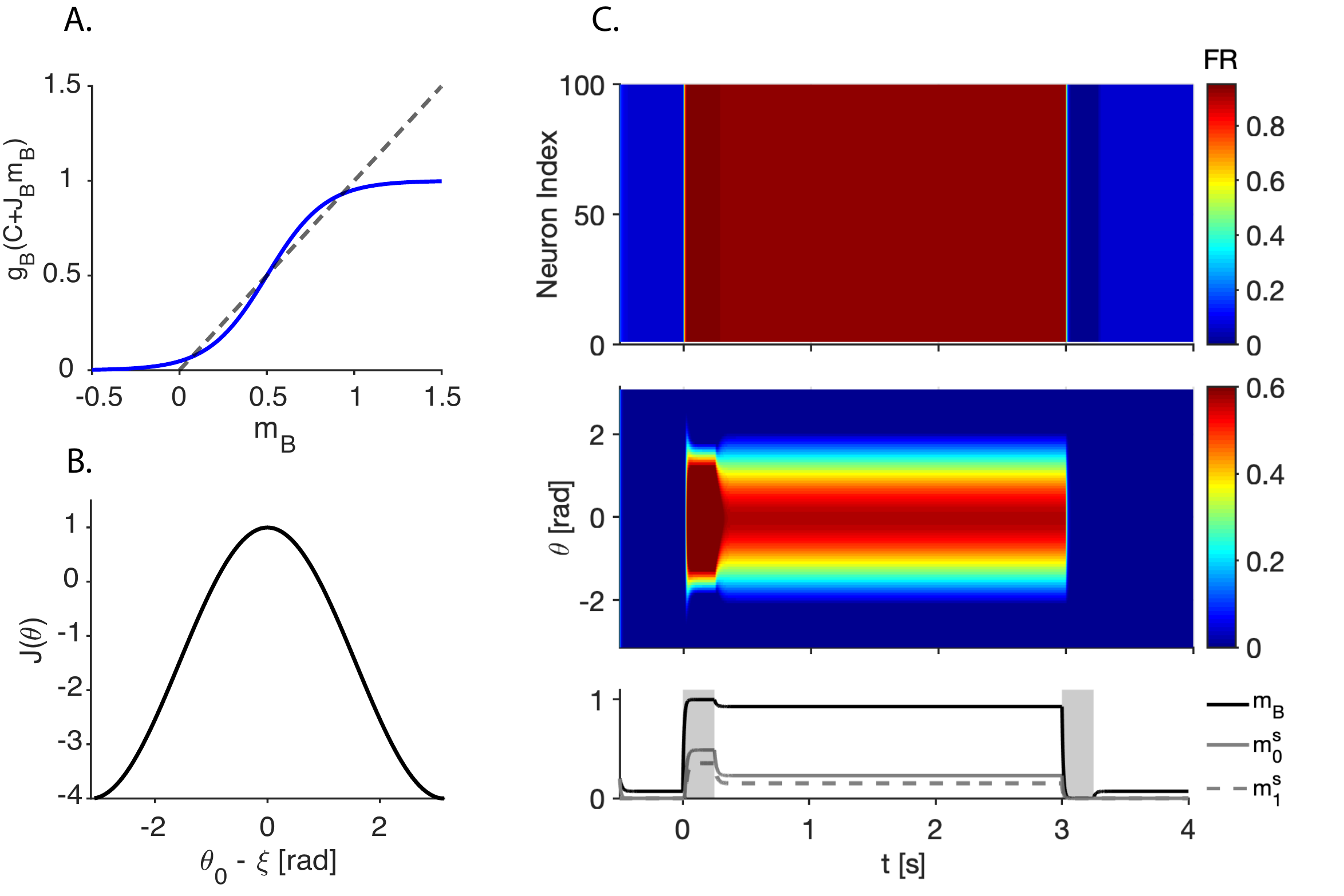}
\end{center}
\caption[Two-module model.]{
{\bf Two-module model} \textbf{A.} Graphic solution of equation (\ref{BistableFP}). When the function $g_B(x)$ is sigmoidal there can be multiple solutions. \textbf{B.} Connectivity profile of neuron with PD=0  \textbf{C.} Dynamics of the network when subjected to a 0.3 s excitatory pulse at $t = 0$ s centered at $\theta = 0$ and a 0.3 ms inhibitory pulse at $t = 3$ s. The values of $m_B$ and $m_S(\theta)$ are given in the color code. Bottom: $m_B$ (solid black line) and the zero- and first- order Fourier components of $m_S$ as a function of time (solid and dashed gray lines, respectively).  
}
\centering
\label{fig:DivConqDyn}
\end{figure}

There are many ways to implement a bistable, non-selective network; here, we depict a simple way, in which bistability rises from the nonlinearity of the neuronal transfer function (TF). In this network neurons are not direction selective, i.e all the neurons receive the same external input when the cue is present regardless of its direction. Within the network, the input each neuron receives from the network is the weighted sum of the activities of all incoming synapses from the other neurons; we denote these activities by $m_B^i$ ($i = 1,2,...,N$). The weight of the connection between two neurons in the bistable module with indexes $i$ and $j$ is $J_{ij} = J_B/N$, therefore the recurrent input to each neuron is $J_B\frac{1}{N}\sum_{j=1}^N m_B^j$. In addition, each neuron receives a background input $C$ and an external input $I_{ext}$ which are identical for all the neurons. The outputs of the neurons are their firing rates, $r_B^i$, which are linked to their inputs via the neurons' TF, $g\left(I\right)$:
\begin{equation*}
r_B^i = g_B\left(C+I_{ext}+J_B\frac{1}{N}\sum_{j=1}^N m_B^j\right)
\end{equation*}
The outgoing synapses of neuron $i$ have linear dynamics with its firing rate as input, i.e. $\tau \dot{m}_B^i = -m_B^i + r_B^i$. In the steady state all the synapses have the same activity level $m_B$,  and the self consistent equation for the activity is:
\begin{equation}\label{BistableFP} 
m_B=g_B\left(C+J_Bm_B+I_{ext}\right) 
\end{equation} 
Acknowledging the saturation of the TF by taking $g_B(I)$ to be sigmoidal implies that in the absence of an external input there can exist some values of $C$ for which equation (\ref{BistableFP}) has multiple solutions (figure \ref{fig:DivConqDyn}A). The system would therefore have two stable steady states, and a transient external input can be used to shift between them. A mechanism of this nature will be discussed in a later section; in addition, it will be demonstrated that it is also possible to implement a bistability of homogeneous states without saturating transfer functions if we consider the contribution of inhibitory neurons in the network.

A network that displays either a low homogeneous state or a bump state has already been studied by Ben-Yishai et al. in \cite{Ben-Yishai1995} with the \enquote{ring model} framework. Within this model arises a case with the dynamics:
\begin{equation}\label{m_BY} 
\tau  \dot{m_S}(\theta,t)=-m_S(\theta,t)+g_S\left(C+\frac{1}{2\pi}\int_{2\pi}{J(\theta-\xi)m_S(\xi,t)\mathrm{d}\xi} + I_{inp}\right) 
\end{equation} 
where $C$ is a background input, the connectivity $J$ is taken to be $J(\theta-\xi) = J_0 + J_1\cos(\theta-\xi)$ (figure \ref{fig:DivConqDyn}B) and the transfer function of the selective module, $g_S(I)$, is the threshold linear function, $g_S(I) = \max(I,0)$. In the absence of an external input, if $C<0$ then the only possible steady state is a homogeneous steady state in which $m_S(\theta) = 0$. If $C$ is positive and $J_1>2$ then there is again only one steady state, but this time it is shaped like a \enquote{bump} of activity. Because of the symmetry of the network, the center of this bump is determined by asymmetries in the initial conditions (which can be induced by a transient input). The input $I_{inp}$, in our case, is the sum of an external input $I_{ext}$ and an input proportional to the activity in the bistable module, $J_{SB}\cdot m_B$. When $I_{ext} = 0$, therefore, this module will be in a homogeneous state if $C+J_{SB}\cdot m_B<0$ and in a bump state if $C+J_{SB}\cdot m_B>0$.

Figure \ref{fig:DivConqDyn}C displays the dynamics of this implementation. During the fixation period ($t<0$) both modules do not receive any input. At $t = 0$ s the bistable module receives a 300 ms pulse of homogeneous excitation, while the selective module receive a small, tuned input, centered at $\theta = 0$. At $t = 0.3$ s the inputs are removed, the bistable module relaxes to the up state and the selective module remains in a bump state. At $t = 3$ s a 300 ms inhibitory homogeneous reset input is given to both modules, shifting the bistable module back to its down state and so extinguishing the bump.  

In this mechanism there is a functional segregation between the notion that information must be stored in the memory and the information itself: the selective module is the one that holds the information about the cue location, and the bistable module only prevents it from fading before the task is complete, without access to the information itself. An advantage of this scheme is that the output of the bistable module can be relayed to other modules that hold more information about the cue - shape, color, etc. - and maintain them in memory too if the task so requires. In addition, it could relate to the existence of non selective (\enquote{omni directional}) neurons in the PFC  \cite{Funahashi1989}. A shortcoming, however, is that in this implementation the bistable module can only be switched off by inhibition and thus cannot explain the rise in activity sometimes seen in PFC neurons during the response period \cite{Funahashi1989, Funahashi1990, Funahashi1991}. In the next sections we show a mechanism in which persistence and selectivity occur in a single network, and the switch from the bump state back to the homogeneous state can indeed be done by homogeneous excitation.
\subsection*{Persistence and selectivity within a single network}\label{ssect:CompClever}

\begin{figure}[!h]
\begin{center}
\includegraphics[width=5.5in]{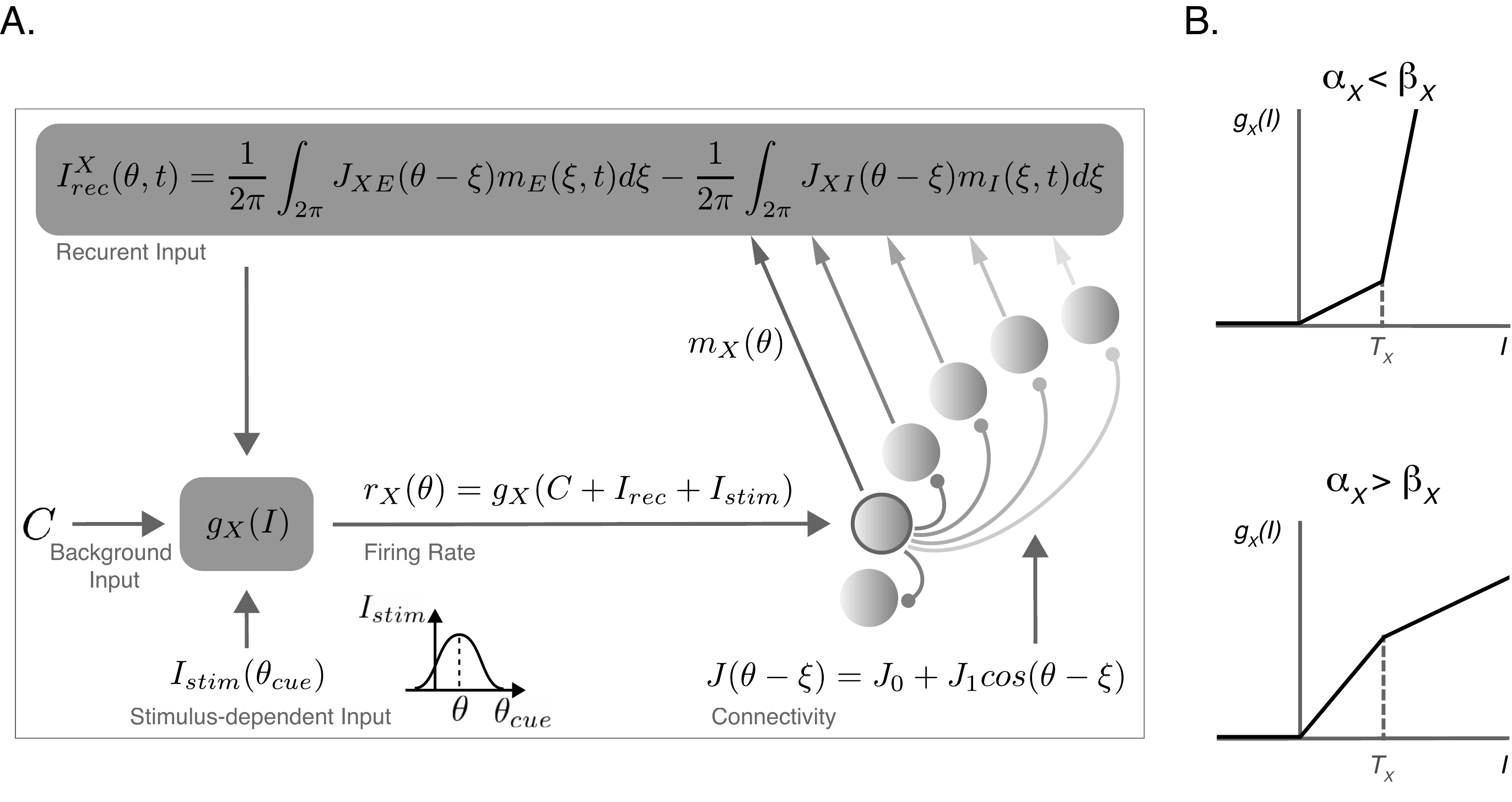}
\end{center}
\caption[Rate model diagram.]{
{\bf Rate model diagram} \textbf{A.} A diagram of the rate model. \textbf{B.} Transfer functions that are considered in this work. For $0<I<T_X$ the slope of $g_X(I)$ is $\alpha_X$, and for $T_X<I$ the slope is $\beta_X$. The $\alpha_X<\beta_X$ case is termed \enquote{expansive} nonlinearity (top) and the $\alpha_X>\beta_X$ is termed \enquote{compressive} nonlinearity (bottom).
}
\centering
\label{fig:ModDiag}
\end{figure}

Neuronal TFs are typically nonlinear and, in the presence of a noisy input, have a sigmoidal shape \cite{arsiero2007}. To analyze the contribution of neuronal nonlinearities to the existence of persistence and selectivity in a single network we implement nonlinear neuronal TFs in a model similar to the above mentioned \enquote{ring model}  \cite{Ben-Yishai1995}. In our model, the network consists of $N_E$ excitatory and $N_I$ inhibitory neurons, where each neuron is labeled according to its \enquote{preferred} direction (PD) -  the stimulus direction for which the input to this neuron is maximal. The preferred directions $\theta$ are evenly distributed on the segment $[-\pi, \pi]$.  The inputs to the neuron are a background input $C$, the stimulus dependent input $I_{stim}$ and a recurrent input from the network $I_{rec}$ (figure \ref{fig:ModDiag}A). The output of the neuron is its firing rate $r_X$, which is determined by the neuron's TF, $g_X(I)$ ($X\in\{E,I\}$). To study the effects of nonlinearities in the TFs we take $g_X(I)$ to be piecewise linear: when the input is below some value $T_X$ the slope is $\alpha_X$; above this value the slope is $\beta_X$. This framework allows us to study different aspects of the TF - an increasing slope in $g_X(I)$ approximates the typical power-law acceleration of neuronal TFs close to the threshold (figure \ref{fig:ModDiag}B, top), while a decreasing slope represents the saturation of the neuron's firing rate at high input intensities (figure \ref{fig:ModDiag}B, bottom). The recurrent input is the weighted sum over the synaptic rates $m_X(\theta)$. The connectivity has a \enquote{Mexican-hat} shape: the strength of the synaptic connection between a presynaptic neuron with PD $\xi$ in population $X$  and a postsynaptic neuron with PD $\theta$ is $J_X(\theta-\xi) = J_X^0 + J_X^1\cos(\theta-\xi)$. The full details of the model are in the Methods section.
%
%
%
%

{\bf Reduction to one population.} We start with a simple case where the excitatory and inhibitory neurons' TFs and time constants are identical, i.e. $g_E(I) = g_I(I) \triangleq g(I)$ and $\tau_E = \tau_I \triangleq \tau$. This enables us to describe the activity profiles of both the excitatory and the inhibitory populations by a single variable $m(\theta)$, since the TFs and inputs are identical in both populations. We take $T_E = T_I = 1$,  $\alpha_E = \alpha_I = 1$ and $\beta_E = \beta_I = \beta$, so that the scale of the rate is set by $\beta$. A model in which $\beta>1$ reflects the TF acceleration whereas a model with $0<\beta<1$ displays the TF saturation; we term the case $\beta>1$ as \enquote{expansive} nonlinearity and the case $\beta<1$ as \enquote{compressive} nonlinearity.

{\bf Steady states.} Due to the symmetry of the network we can consider without loss of generality only steady state profiles that are symmetric around $\theta = 0$. The fixed point (FP) equation is therefore:  
\begin{equation}\label{Res1popFP} 
m(\theta)=g\left(C+J_0m_0+J_1m_1\cos(\theta)\right) 
\end{equation}   
where $J_k = J^E_k - J^I_k$ and $m_k = \frac{1}{2\pi}\int_{2\pi}{m(\theta)\cos(k\theta)\mathrm{d}\theta}$. 

In the case of the homogeneous FP we have $m(\theta) = m_0$ ($m_1 = 0$), and the steady state equation is reduced to equation (\ref{BistableFP}). This equation can be rewritten as
\begin{equation}\label{ResHomFP} 
I^*-C = J_0g(I^*)
\end{equation} 
where $I^* = C+J_0m_0$ is the total input to the neuron at the steady state. Similarly to the case studied in \cite{Ben-Yishai1995}, the stability of this solution depends on $g'(I^*)$, the slope of $g(I)$ at the point $I^*$. If $J_0g'(I^*)>1$ the network is \enquote{rate-unstable}, meaning it is unstable under small homogeneous perturbations;  if $J_1g'(I^*)>2$ the network is \enquote{spatially unstable} and small deformations of the profile can drive it away from the FP.

\begin{figure}[!b]
\begin{center}
\includegraphics[width=6in]{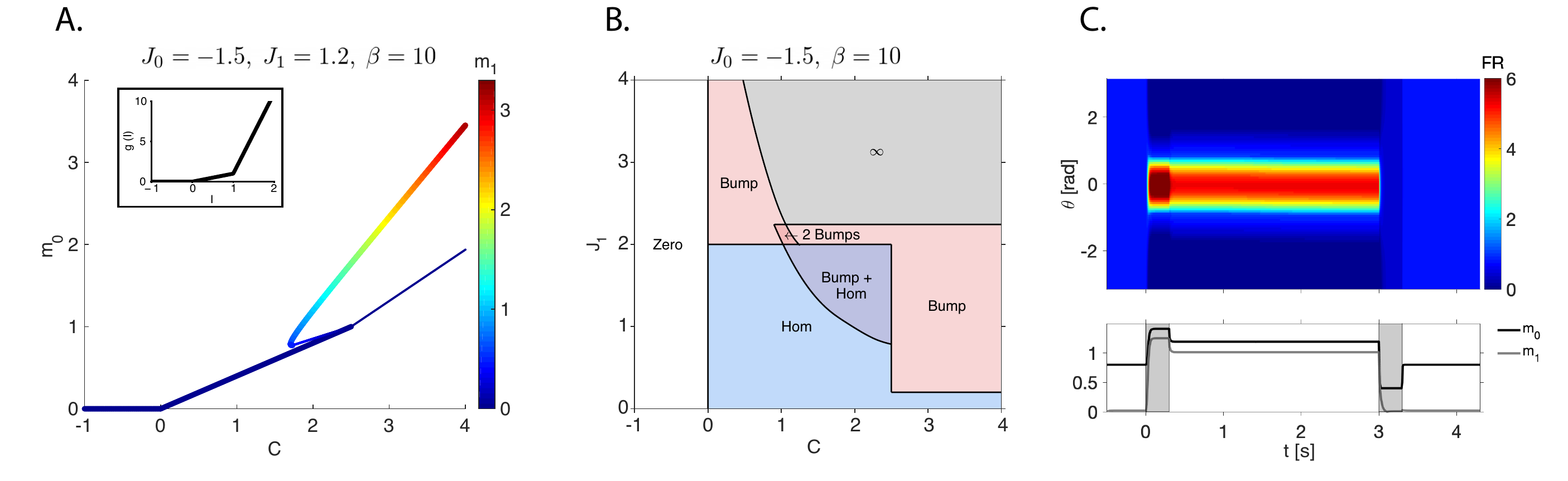}
\end{center}
\caption[Expansive nonlinearity model]{
{\bf Expansive nonlinearity ($\beta>1$).} \textbf{A.} Bifurcation diagram of $m_0-C$ ($m_1$ in color code) for $J_0 = -1.5$, $J_1 = 1.2$ and $\beta = 10$. Thick: stable state. Thin: unstable state. Note the regime of $C$ where a stable state with $m_1 = 0$ coexists with a stable state with $m_1>0$. Inset: the neuronal transfer function $g(I)$. \textbf{B.} Phase diagram in the $C-J_1$ plane for $J_0 = -1.5$ and $\beta = 10$. \textbf{C.} Dynamics of the network in response to a 0.3 s tuned excitatory pulse at $t = 0$ centered at $\theta = 0$, and a 0.3 s untuned inhibitory pulse at $t = 3$. Top: $m(\theta)$ (color code) as a function of time. Bottom: Fourier components of the activity profile as a function of time. Parameters are as in \textbf{A} and $C = 2$. 
}
\centering
\label{fig:1popExpNonLin}
\end{figure}

An immediate result of this stability analysis is that, since the slope $g(I)$ is not constant, it is possible for the homogeneous state to be stable in one regime of $I^*$ but unstable in another. In the expansive nonlinearity model ($\beta>1$), for example, if ${J_0<1}$ and ${2>J_1>2/\beta}$ then the homogeneous state is spatially stable if ${I^*<1}$ but unstable if ${I^*>1}$. In this case, an increase in $C$ can cause the destabilization of the homogeneous state and the emergence of a bump in a subcritical pitchfork bifurcation (figure \ref{fig:1popExpNonLin}A). This bifurcation creates a regime of $C$ where the homogeneous state co-exists with the bump.
As can be seen from the phase diagram in figure \ref{fig:1popExpNonLin}B, when $C$ is small, an increase in $J_1$ follows the scenario described in \cite{Ben-Yishai1995} - the homogeneous state loses stability when $J_1 = 2$ and is replaced by a bump. A similar process occurs upon $J_1$ increase when $C$ is large, except that in this case the homogeneous state is in the upper branch of the TF (where $g'(I) = \beta$) and therefore it loses stability at $J_1 = 2/\beta$. The transition between these two regimes gives rise to bistability: the lower homogeneous state coexists with a bump that rises from the spatial instability of the upper homogeneous state. 
%

When $C$ is in the bistable regime, it is possible to shift from the homogeneous state to the bump state (\enquote{switch-on}) and back (\enquote{switch-off}) by transient inputs. A simple switch-on mechanism can use the steady state properties of the network. When on the homogeneous branch in the bistable regime of the bifurcation diagram (figure \ref{fig:1popExpNonLin}A), an increase in $C$ will lead the network to a regime where the only stable state is a bump. A homogeneous input will therefore cause the network to converge to a bump state, and a slight tuning of the input can suffice in determining the center of the bump. When the input is removed, the network returns along the stable bump branch back to the bistable regime. Similarly, a decrease in $C$ shifts the network to a regime where only a homogeneous state exists, hence the bump can be switched off by inhibition. Figure \ref{fig:1popExpNonLin}C shows the response of the network to a 300 ms tuned excitatory input centered at $\theta = 0$ at $t = 0$ s, and a 300 ms homogeneous inhibitory input at ${t = 3\,\textrm{s}}$ (model parameters are as in figure \ref{fig:1popExpNonLin}A with $C = 2$). Following the excitatory input the network shifts from a homogeneous state to a bump state and $m_1$ obtains a positive value. The transient homogeneous input at $t = 3$ s shifts the network back to the homogeneous state.

\begin{figure}[!b]
\begin{center}
\includegraphics[width=6.1in]{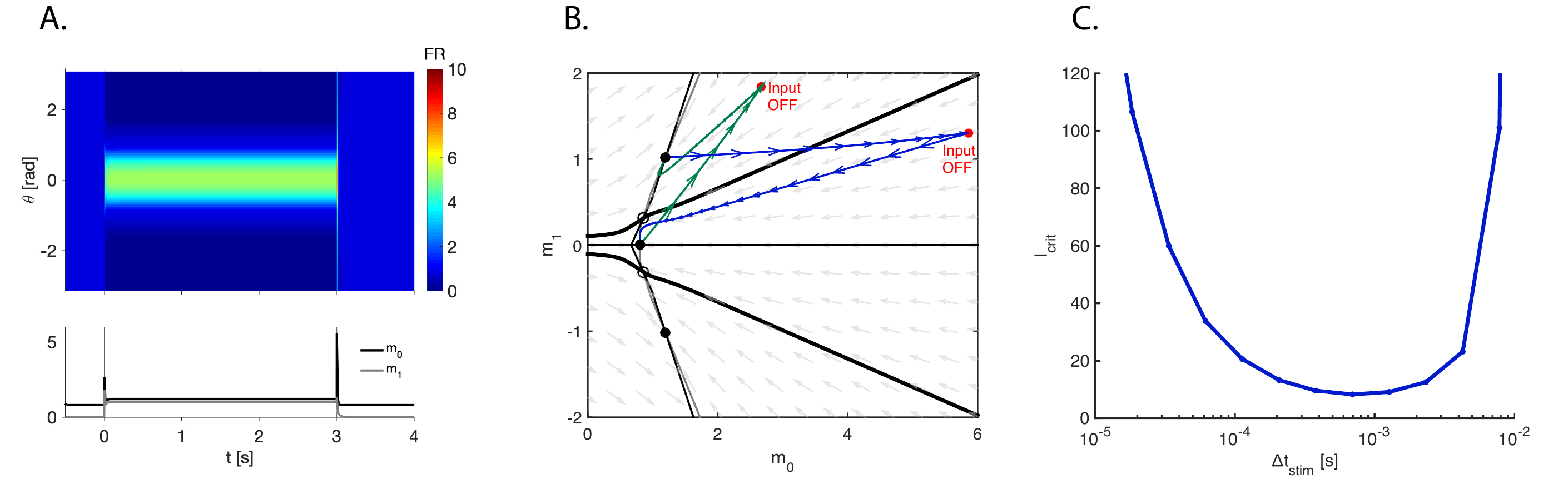}
\end{center}
\caption[Dynamics of the expansive nonlinearity model]{
{\bf Dynamics of the expansive nonlinearity model.} \textbf{A.} Dynamics of the network in response to a 3 ms tuned excitatory pulse at $t = 0$ centered at $\theta = 0$, and a 0.5 ms homogeneous excitatory pulse at $t = 3$.  Parameters are similar to figure \ref{fig:1popExpNonLin}C. \textbf{B.} Dynamics of $m_0$ and $m_1$ from \textbf{A} plotted on the $m_0-m_1$ phase plane (see text). Green: trajectory during switch on ($0<t<3$ ms). Blue: trajectory during switch off (${3<t<3.0005\,\textrm{s}}$). {\textbf{C.} Minimal switch-off input amplitude, $I_{crit}$, plotted against the duration of that input, $\Delta t$. When the homogeneous excitatory input is too long switch off can not be done. dt=0.01 ms; $\tau$=10 ms.}}
\centering
\label{fig:1popExpNonLinDyn}
\end{figure}

Alternatively, the switching can also be performed by transient inputs on a time scale smaller than that of the synapses; in this case, the mechanisms rely on the dynamics rather than on the attractors. Figure \ref{fig:1popExpNonLinDyn}A displays the response of the network to a 3 ms tuned input at $t = 0$ s and a 0.5 ms homogeneous excitatory input at $t = 3$ s. As can be seen, the bump is switched on by the tuned input, and is switched off by the homogeneous input which, in contrast to the prolonged switch-off input (figure \ref{fig:1popExpNonLin}C), must be excitatory to perform the switch off. We analyze these switching mechanisms from a dynamical point of view by studying the dynamics of the order parameters $m_0$ and $m_1$ (equation \ref{Res1popFP}) following a symmetric input, with the initial conditions set to the either the bump or the homogeneous FP. Figure \ref{fig:1popExpNonLinDyn}B displays the $m_0-m_1$ phase plane; since the initial conditions and the input are symmetric around $\theta = 0$ we need not worry about the sine term, $m_2$ (see Methods). The thin black and gray lines are the $m_1$ and $m_0$ nullclines, respectively, and filled and empty circles represent the stable and unstable FPs, respectively. There is co-existence of a stable homogeneous FP and stable bump FPs: the FP on the $m_0$ axis corresponds to the homogeneous state and the FP on the upper (resp. lower) half plane corresponds to the bumps centered at $\theta = 0$ ($\theta = \pi$). The thick black lines are the stable manifolds of the unstable FPs, and they also separate the basins of attraction (BOAs) of the stable homogeneous state and the stable bump state. The green trajectory depicts $m_0$ and $m_1$ from figure \ref{fig:1popExpNonLinDyn}A following the switch-on input. From this trajectory it can be seen that the input duration need only be long enough for the trajectory to cross the separatrix: as long as this input is withdrawn with the trajectory inside the BOA of the bump state (red dot) the switch-on will occur. Similarly, the blue trajectory shows $m_0$ and $m_1$ from figure \ref{fig:1popExpNonLinDyn}A following the switch-off input. This time the input is withdrawn inside the BOA of the homogeneous state, and the trajectory terminates at the homogeneous FP.

Unlike the switch-on, here the switch-off input can not be prolonged: if the input is turned off too late, the trajectory would turn towards the positive direction of the $m_1$ axis and cross the separatrix back into the BOA of the bump FP. In this case it would terminate back at the non-homogeneous FP and the bump would not be switched off. But even when the input is sufficiently brief the input must still be strong enough for the trajectory to cross the separatrix into the BOA of the homogeneous FP. For every stimulus duration $\Delta t$ there is therefore a minimal input intensity $I_{crit}$ for successful switch-off; this relation is plotted in figure \ref{fig:1popExpNonLinDyn}C. As $\Delta t$ becomes smaller, a larger amplitude of the input is needed for the trajectory to enter the BOA of the homogeneous state. When $\Delta t$ is enlarged $I_{crit}$ decreases, but then it increases again since the trajectory must go deep enough into the BOA of the homogeneous state to remain inside it during $\Delta t$. When $\Delta t$ exceeds the time scale of the synaptic time constant the value of $I_{crit}$ grows to infinity and switch-off becomes impossible, since $\Delta t$ becomes large enough to allow the system to converge to the bump fixed point which exists at large $C$. 

\begin{figure}[!b]
\begin{center}
\includegraphics[width=4in]{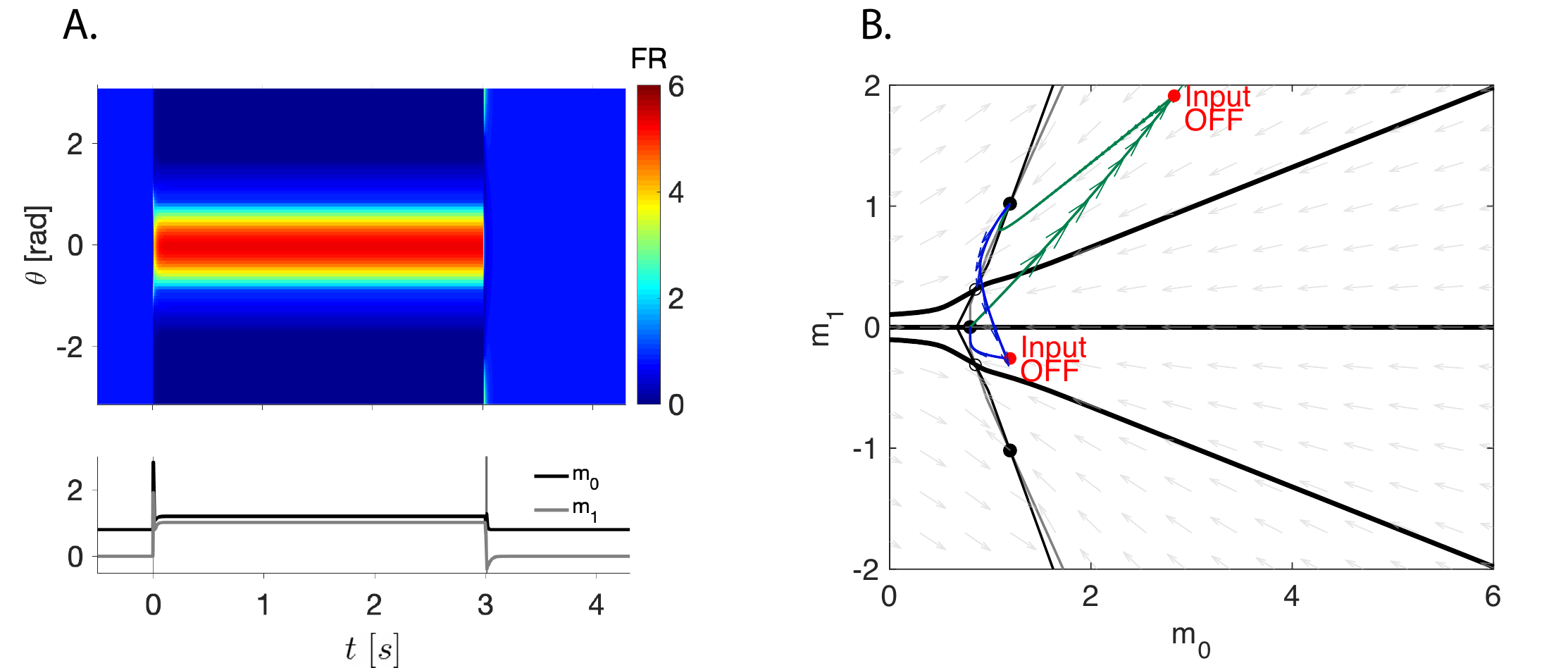}
\end{center}
\caption[Expansive nonlinearity, switch of with anti-phase excitation]{
{\bf Expansive nonlinearity, switch of with anti-phase excitation.} \textbf{A and B}. Same as in figure \ref{fig:1popExpNonLinDyn}\textbf{A} and \textbf{B}, except that the switch off input is a 12 ms tuned excitatory pulse, centered at $\theta = \pi$.
}
\centering
\label{fig:1popExpNonLinSwAntPhs}
\end{figure}

One way to extend the regime of $\Delta t$ in which switch-off is possible is by inserting a tuned (rather than a homogeneous) excitatory transient input, centered at the direction opposite to the bump center. Figure \ref{fig:1popExpNonLinSwAntPhs}A shows the response of the network when the switch-on input is as before and the switch-off input is a 12 ms cosine-shaped excitation centered at $\theta = \pi$ ($I_{stim}(\theta) = 10\cdot\left( 1+\cos(\theta-\pi)\right)$). During the switch-off period the network starts to develop a bump around $\theta = \pi$, but as the input is removed it relaxes back to the homogeneous state. This is also shown by the blue trajectory in figure \ref{fig:1popExpNonLinSwAntPhs}B: before the input is removed the trajectory is drawn to the lower half-plane, but the input is turned off while the trajectory is within the BOA of the homogeneous FP. Although this scheme enables switch-off with a longer time scale than the homogeneous switch-off input it still has certain restrictions. First, if the input is too long then the sine Fourier component $m_2 = \frac{1}{2\pi}\int_{2\pi}m(\theta)\sin(\theta)\mathrm{d}\theta$ becomes larger and the trajectory leaves the BOA of the homogeneous FP  in the $m_2$ direction. Second, for the same reason, this switch-off scheme is sensitive to the location of the switch-off input: when it is perfectly centered at $\theta = \pi$ then only $m_1$ receives input, but any deviation will cause $m_2$ to receive input as well and possibly drive the trajectory further  away from the $m_0$ axis. We therefore conclude that in this model, any switch-off mechanism that is based on excitation must be fast relative to the synaptic time scale.

\begin{figure}[!b]
\begin{center}
\includegraphics[width=6.1in]{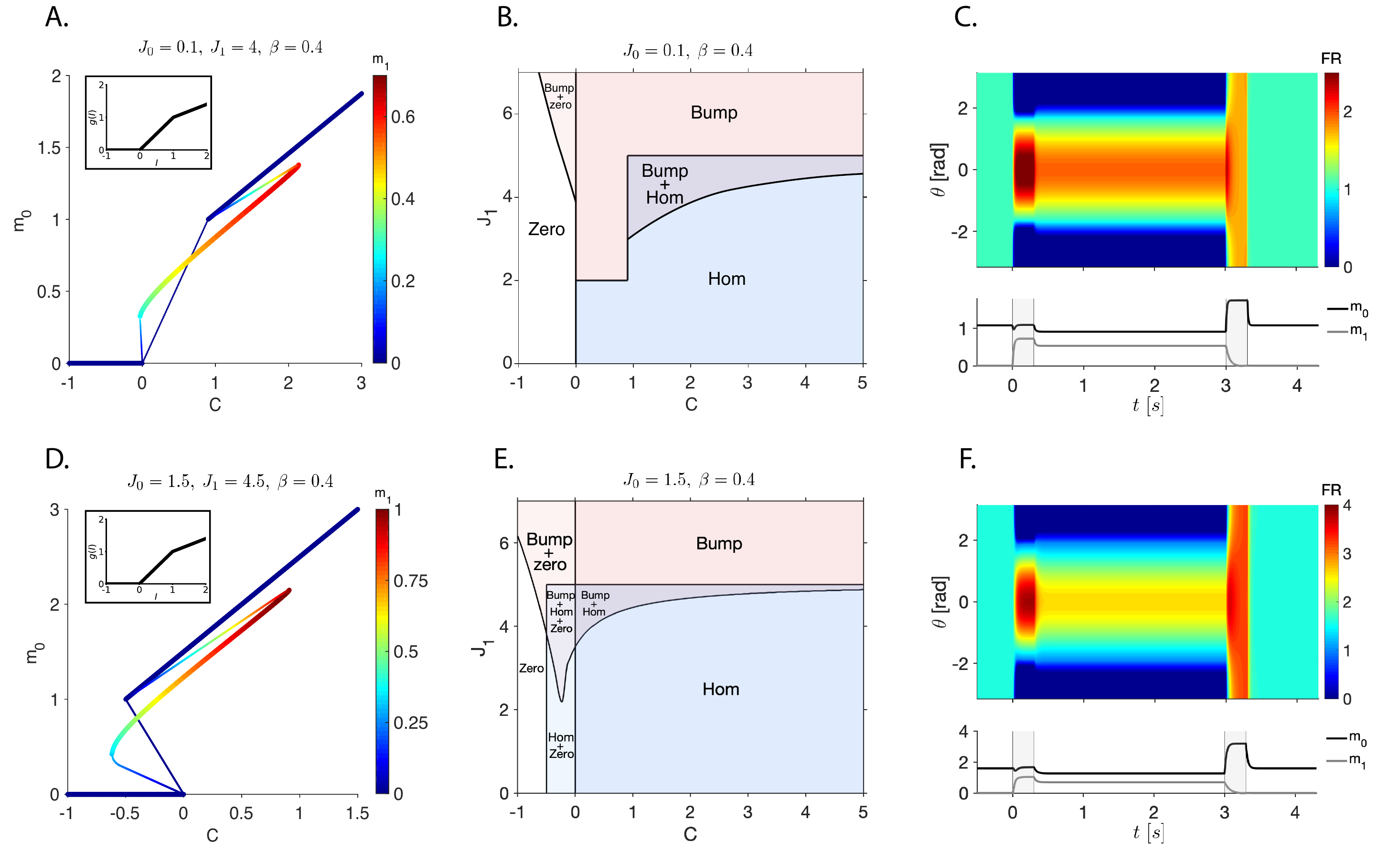}
\end{center}
\caption[Compressive nonlinearity model]{
{\bf Compressive nonlinearity model ($\beta<1$).} \textbf{A-F}: $\beta = 0.4$. \textbf{A} and \textbf{D}. graphics are as in figure \ref{fig:1popExpNonLin} \textbf{A} with $J_0 = 0.1$, $J_1 = 4$ (in \textbf{A}) and $J_0 = 1.5$, $J_1 = 4.5$ (in \textbf{D}). In the case ${J_0<1}$ (\textbf{A}) the unstable homogeneous state is only \textit{spatially} unstable while in the case ${J_0>1}$ (\textbf{D}) the unstable homogeneous state is both spatially- and rate-unstable. \textbf{B} and \textbf{E}. Phase diagram in the $J_1-C$ plane for $J_0 = 0.1$ and $J_0 = 1.5$, respectively. In \textbf{E}, note the existence of a regime where two stable homogeneous states exist. In the low homogeneous state $m_0$ is zero. \textbf{C} and \textbf{F}. Dynamics of the network and of the order parameters in response to a 0.3 s tuned input at ${t = 0\,\textrm{s}}$ centered at $\theta = 0$, and a 0.3 s homogeneous excitatory pulse at $t = 3$ s. $C=1.1$ and 0.1, respectively. Here, in both cases switch-off is possible with a prolonged homogeneous pulse.
}
\centering
\label{fig:1popCompNonLin}
\end{figure}

In terms of switching mechanisms, the compressive nonlinearity (${\beta<1}$) case is the opposite of the expansive nonlinearity case. In the latter,  switch-off with a prolonged homogeneous input is possible only if the input is inhibitory; this is because the increase in TF slope causes the homogeneous state to lose stability as the total input is \textit{increased}, and therefore when $C$ is large the only attractor is a bump. In the compressive nonlinearity case the situation is inverted:  if ${J_0<1/\beta}$ and ${2<J_1<2/\beta}$ then the homogeneous state that is on the upper branch of the TF (where ${g'(I) = \beta}$) is stable whereas the homogeneous state on the lower branch is unstable. The spatial instability, again, leads to the appearance of a bump in a sub-critical pitchfork bifurcation. Here, in contrast to the expansive nonlinearity model, the bump does not exist when $C$ is large (figure \ref{fig:1popCompNonLin}A and D). 

Interestingly, in the compressive nonlinearity model we can distinguish between a case where the homogeneous state on the lower branch is only spatially unstable and a case where it is also rate-unstable. In the former, which occurs if $J_0<1$, there is a regime of $C$ ($0<C<1-J_0$) in which no stable homogeneous solution exists (figure \ref{fig:1popCompNonLin}A). In the latter, in which $J_0>1$, there is no such regime; instead, for $1-J_0<C<0$ there is bistability of homogeneous solutions in addition to the stable bump (figure \ref{fig:1popCompNonLin}D). The mechanism underlying this bistablity has been mentioned in the functionally-segregated network: since the homogeneous state is determined by the solution to equation (\ref {ResHomFP}), in the case $J_0>1$ there can be 3 solutions to this equation. Note also that in both cases, and in contrast to the expansive nonlinearity model, the average rate of the bump is lower than the rate in the homogeneous state. 

As in the expansive nonlinearity model, here the network can be switched on by a tuned input; the difference is that now the only state when $C$ is large is a homogeneous state, and therefore the tuning of the input can not be too small in order to switch the bump on. Another substantial difference, that stems from the same reason, is that the network can be switched off by a prolonged {\it excitatory} homogeneous input. In figure \ref{fig:1popCompNonLin}C and F we show the dynamics of the network when subjected to a 300 ms tuned input at $t = 0$ s, and a 300 ms homogeneous excitatory input at $t = 3$ s. After the tuned input is removed the network stays in a bump state, until the transient homogeneous input switches the network back to the homogeneous state. The bump is wider than in the expansive nonlinearity model since the difference in firing rates between neurons becomes smaller around the center of the bump (rather than larger in the expansive nonlinearity model).  

\begin{figure}[!t]
\begin{center}
\includegraphics[width=6in]{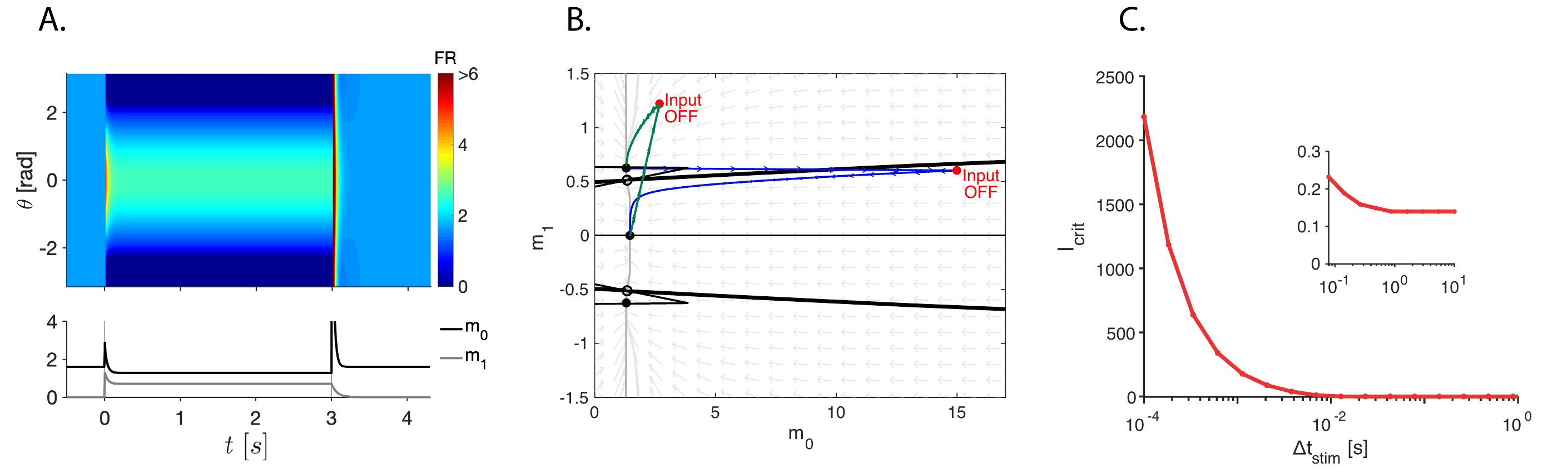}
\end{center}
\caption[Switching between states in the compressive nonlinearity model]{
{\bf Switching between states in the compressive nonlinearity model} Parameters are similar to figure \ref{fig:1popCompNonLin}A. \textbf{A.}  Dynamics of the network and order parameters in response to a 3 ms tuned input at $t = 0$ s centered at $\theta = 0$, and a 0.5 ms homogeneous input at $t = 3$ s. \textbf{B.} Dynamics of $m_0$ and $m_1$ from A plotted on the $m_0-m_1$ phase plane. Green: response to the switch-on input. Blue: response to switch-off input. \textbf{C.} Minimal switch-off input amplitude, $I_{crit}$, plotted against the duration of that input, $\Delta t$. Inset: close up on the right end of the curve ($\Delta t>0.1$). $\tau$=10 ms.
}
\centering
\label{fig:1popCompNonLinDyn}
\end{figure}

Switching can also be performed by transient inputs with durations smaller than the synaptic time scale, as can be seen in figure \ref{fig:1popCompNonLinDyn}A, in which the switch-on input is a 3 ms tuned input and the switch-off input is a 0.5 ms homogeneous input. In figure \ref{fig:1popCompNonLinDyn}B ,the trajectories of the switch-on and switch-off are plotted on the $m_0-m_1$ phase plane in green and blue, respectively. As for the switch-off, a more prolonged input would result in the network being effected by the longer time scale synaptic dynamics, with the trajectory bending towards the $m_0$-axis, deeper into the BOA of the homogeneous state. However, as can be seen from the phase plane, since the separatrix itself is pulling away from the $m_0$ axis as $m_0$ increases, this bending of the trajectory is unnecessary for the trajectory to penetrate the BOA of the homogeneous state. Like in the expansive nonlinearity model, there is a minimal input intensity $I_{crit}$ for every input duration $\Delta t$ in order for the trajectory to cross the separatrix (figure \ref{fig:1popCompNonLinDyn}C). The difference is that here when $\Delta t$ is large the switch-off can still be performed and $I_{crit}$ does not diverge as a function of $\Delta t$; rather, it converges to a finite value (figure \ref{fig:1popCompNonLinDyn}C, inset).

Another comment about the compressive nonlinearity model is that, as can be seen from figure \ref{fig:1popCompNonLin}B and E, the bistable regime of $C$ is enlarged to an infinite size when $J_1$ approaches ${2/\beta}$. This enlargement, however, implies that the BOA of the homogeneous state becomes smaller, when eventually, when ${J_1 = 2/\beta}$, the unstable branch of the bump (figure \ref{fig:1popCompNonLin}A and D) coalesces with the stable homogeneous branch and destabilizes it. Another implication is that when the bistable regime of $C$ grows, a larger increase in background input is needed to switch off the bump. There is therefore a trade-off: the robustness of the bump and size of the bistable regime come on the expense of the robustness of the homogeneous state and the switch-off.
  
As a step towards generalization of these models, let us define two functions, ${F_0 \triangleq J_0\cdot g(I)}$ and ${F_1 \triangleq J_1\cdot g(I)}$. The mechanism for bistability in the above reduced models originates from the fact that the homogeneous state stability depends on $F'_i(I)$, and $F_0$ and $F_1$ have gains that vary as a function of $I$; upon $I$ increase the homogeneous state can therefore be de-stabilized (in the expansive nonlinearity model) or re-stabilized (in the compressive nonlinearity model), and the bistability arises in the transition between the regimes. Moreover, in the compressive nonlinearity model the deceleration of $F_0$ and $F_1$ enables switch-off with prolonged homogeneous excitation. However, since in this model $F_0$ and $F_1$ are proportional to the neuronal TF, this mechanism explicitly exploits the nonlinearity of the TFs close to their saturation, hence the neurons fire close to their saturation rates both in the homogeneous state and around the center of the bump. In the next section we show that in a more general case the mechanism can be based on decelerating $F_0$ and $F_1$ without involving the saturation of the neuronal TF.   

\subsection*{Effective saturation}

After showing how persistence, selectivity and switch-off by global excitation can arise in a network from the saturation of the neuronal TF, we now demonstrate how the effects of excitatory and inhibitory neurons can be combined to implement this mechanism effectively, without saturating TFs. Here, we consider the same network as in the previous section only without the assumption $g_E(I) = g_I(I)$. The synaptic dynamics in the excitatory and inhibitory populations are therefore not identical, and are described by the variables $m_E(\theta,t)$ and $m_I(\theta,t)$. The total input to the neuron (in the absence of an external drive) is:
\begin{equation}\label{Res2popItot} 
I(\theta) = C + \left(J_E\ast m_E\right)(\theta) - \left(J_I\ast m_I\right)(\theta)
\end{equation}
where the operator $\ast$ is defined here as $(f\ast g)(\theta) = \frac{1}{2\pi}\int_{2\pi}{f(\xi)g(\theta-\xi)\mathrm{d}\xi}$. In the steady state we have $m_X(\theta) = g_X\left(I(\theta)\right)$ ($X\in\{E,I\}$); substituting this in equation (\ref{Res2popItot})  we can write an equation for the input at the steady state, $I^*$:
\begin{equation} \label{Res2popIFP}
I^*(\theta) - C = \frac{1}{2\pi}\int_{2\pi}{F_0\left(I^*(\xi)\right)\mathrm{d}\xi} + \frac{1}{2\pi}\int_{2\pi}{F_1\left(I^*(\xi)\right)\cos(\theta-\xi)\mathrm{d}\xi}
\end{equation}
where $F_i(I) \triangleq J_i^Eg_E(I) - J_i^Ig_I(I)$. Notice here that in the case $g_E(I) = g_I(I)$ these functions match their definition from the previous section. In addition, in the homogeneous steady state, equation (\ref{Res2popIFP}) is reduced to:
\begin{equation}\label{Res2popHomFP} 
I^* - C = F_0(I^*)
\end{equation}   
and the stability conditions for this state are ${F_0'(I^*)<1}$ for rate stability and ${F_1'(I^*)<2}$ for spatial stability (see Methods). Therefore, if $F_i(I)$ are compressive-nonlinear we expect to see bistability and switch-off mechanisms similar to the ones displayed by the reduced model in the compressive nonlinearity case (figure \ref{fig:1popCompNonLin}A and D). Here, however, this does not require that the excitatory nor the inhibitory TFs be saturating. 

\begin{figure}[!b]
\begin{center}
\includegraphics[width=4.5in]{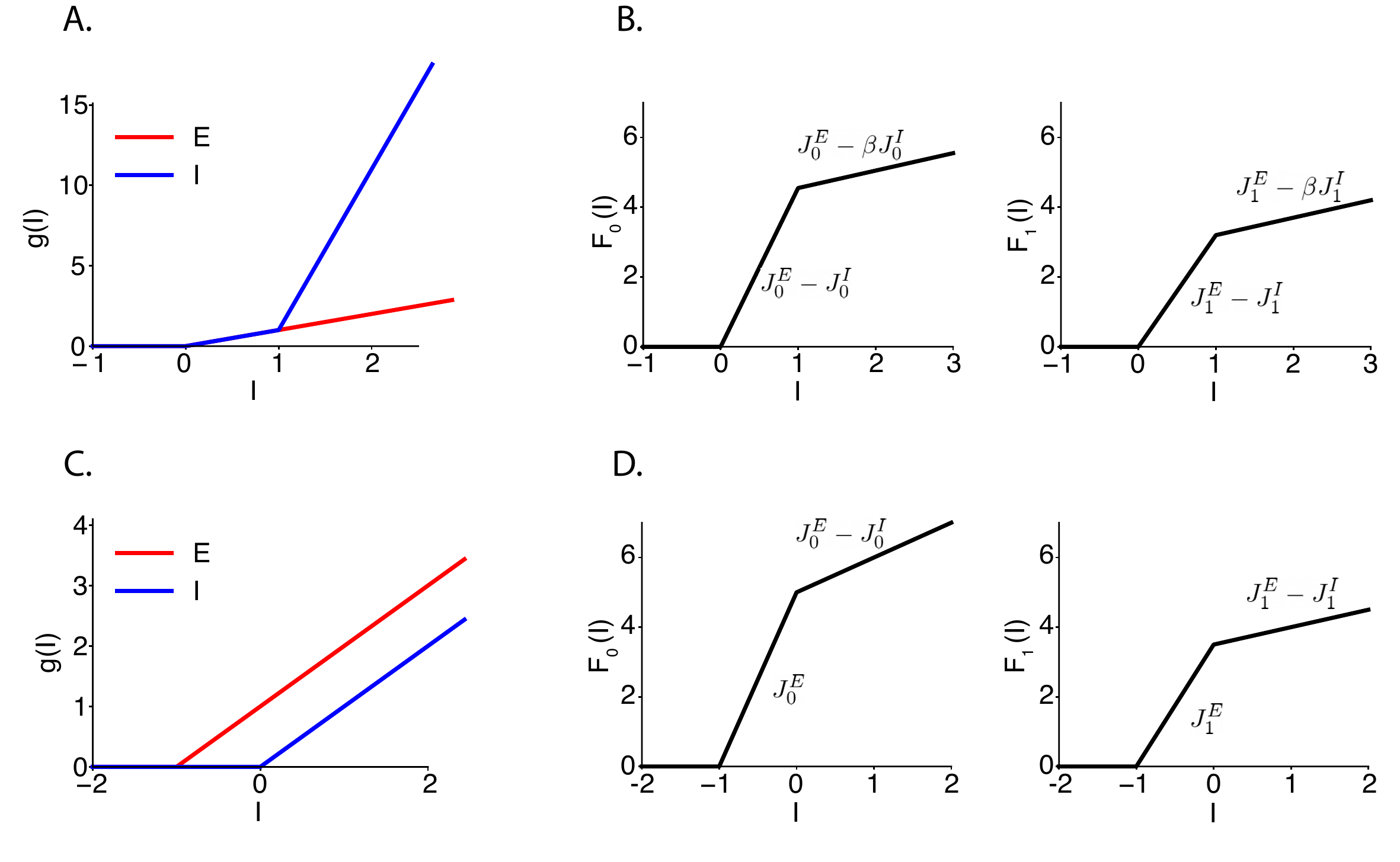}
\end{center}
\caption[]{
{\bf Effective saturation.} \textbf{A.} Excitatory (red) and inhibitory (blue) neurons' TFs in the acceleration difference model. The slope of the inhibitory TF is 1 for ${0<I<1}$ and $\beta$ for $I>1$.  \textbf{B.} Example of $F_0(I)$ (left) and $F_1(I)$ (right) in the acceleration difference model. The slope of the first segment (${0<I<1}$) is $J_i^E-J_i^I$ and of the second segment ($I>1$) is $J_i^E-\beta J_i^I$  ($i = 0,1$). \textbf{C.} Excitatory (blue) and inhibitory (red) neurons' TFs in the threshold difference model. \textbf{D.} Example of $F_0(I)$ (left) and $F_1(I)$ (right) in the threshold difference model. The slope of the first segment (${-1<I<0}$) is $J_i^E$ and of the second segment ($I>0$) is $J_i^E- J_i^I$. 
}
\centering
\label{fig:EffSat}
\end{figure}

An example for such $F_i(I)$ can be obtained by considering the difference in TF acceleration between excitatory and inhibitory neurons. The case where inhibitory TFs have stronger acceleration can be reflected by taking excitatory TFs to be threshold-linear and the inhibitory TFs to be piecewise linear with an increasing slope: when ${I<1}$ the slope is 1, and when ${I>1}$ the slope is $\beta$, where ${\beta>1}$ (figure \ref{fig:EffSat}A). The functions $F_i(I)$, in this case, have a slope $J^E_i-J^I_i$ in the first segment (where ${I<1}$) and ${J^E_i-\beta J^I_i}$ in the second (where ${I>1}$, figure \ref{fig:EffSat}B). Similarly, if we consider a possible difference in thresholds between excitatory and inhibitory neurons we can model it by taking both TFs to be threshold linear, but choose the excitatory TF threshold to be -1 (i.e. $g_I(I) = \max(I,0)$, $g_E(I) = g_I(I+1)$, figure \ref{fig:EffSat}C). In this case the functions $F_i(I)$ will have a slope of $J^E_i$ for ${-1<I<0}$ and a slope of $J^E_i-J^I_i$ for $I>0$ (figure \ref{fig:EffSat}D). For a certain connectivity parameters choice ($J_i ^X$), the slope of the second segment can be smaller than in the first or even negative, while the neuronal transfer functions are strictly increasing.

\begin{figure}[!b]
\begin{center}
\includegraphics[width=5in]{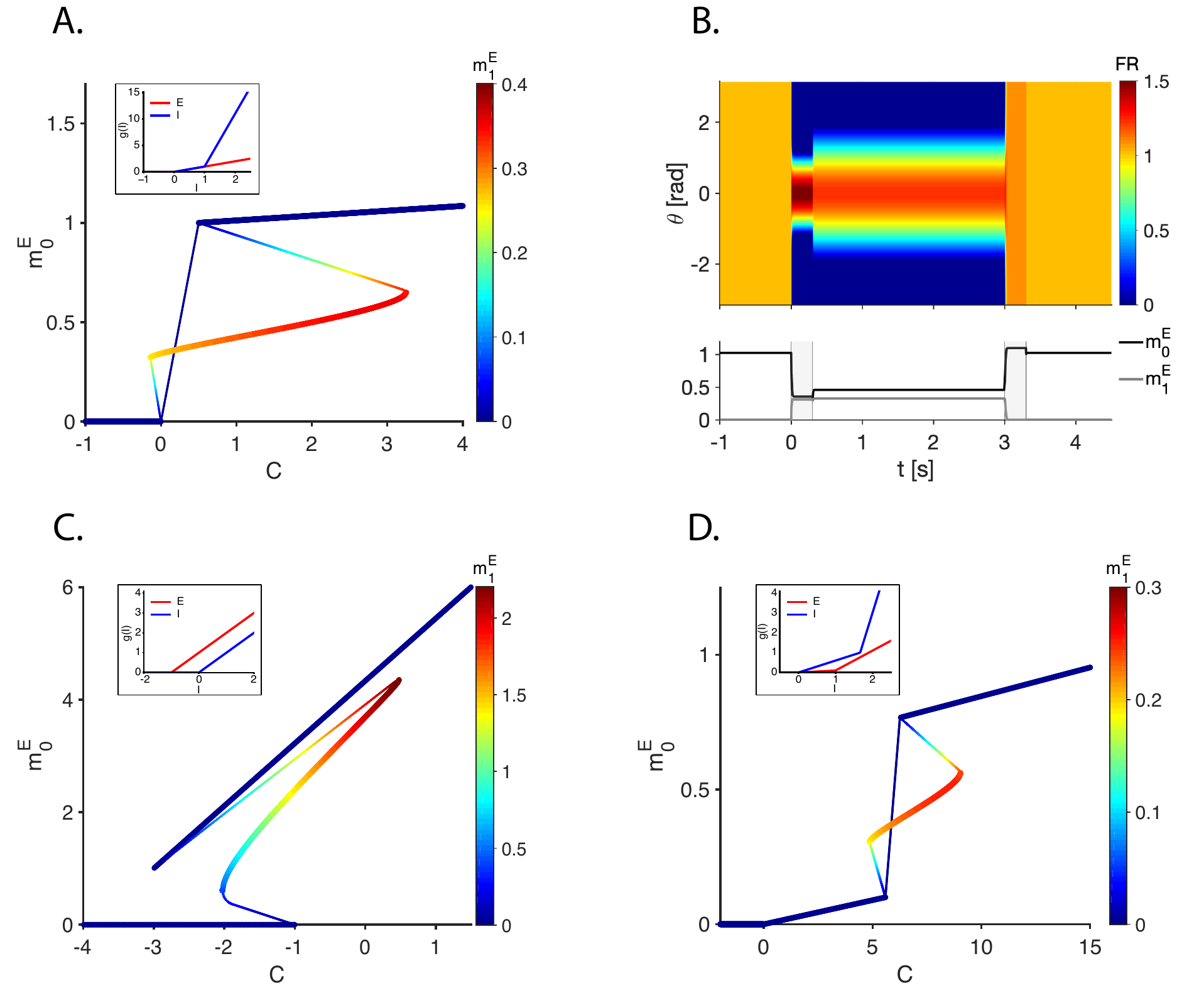}
\end{center}
\caption[Two-population mechanisms.]{
{\bf Two-population mechanisms.} \textbf{A, C and D.} Bifurcation diagrams of $m_0^E-C$ in the 2-population network. Color code: $m_1^E$. Insets: the TFs of the excitatory (red) and the inhibitory (blue) neurons. \textbf{B.} Dynamics of $m_E(\theta)$ (top, color code) and $m_0^E$ and $m_1^E$ (bottom) in response to a 0.3 s tuned input at $t = 0$ s centered at $\theta = 0$ and a 0.3 s homogeneous pulse a $t = 3$ s. Parameters in \textbf{A} and \textbf{B}: $J_{E}^0 = 5$, $J_{I}^0 = 4.5$, $J_{E}^1 = 5$, $J_{I}^1 =  1.1$, $\beta_I = 10$, $C = 1.5$ (in \textbf{B}) Parameters in \textbf{C}: $J_{E}^0 = 3$, $J_{I}^0 =  2.9$, $J_{E}^1 =  3$, $J_{I}^1 =  1.07$ .  Parameters in \textbf{D}: $J_{E}^0 = 5.1$, $J_{I}^0 =  8.5$, $J_{E}^1 =  4.85$, $J_{I}^1 = 1.5$, $\alpha_E = 0.1$, $\beta_E = 1$, $\alpha_I = 0.6$, $\beta_I = 6$, $T_E = 1$, $T_I = 5/3$. 
}
\centering
\label{fig:2Popmech}
\end{figure}

In both these cases, $F_0(I)$ (as well as $F_1(I)$) is piecewise-linear: in the first segment the slope is 0 and $F_0(I) = 0$, in the second segment the slope is $\alpha_F$ and in the last segment the slope is $\beta_F$, where $\alpha_F>\beta_F$. 
As a function of $C$, the line $I ^*-C$ can intersect the curve $F_0(I^*)$ in either the $\beta_F$-, the $\alpha_F$- or the zero-segment, yielding a high, a middle and a low solution, respectively. For the emergence of a bump state it is necessary that the middle solution be spatially unstable; however, since we are interested in a system which can be switched-off by homogeneous excitation, it is also required that the high homogeneous solution be both rate- and spatially-stable.

In the acceleration difference model, the spatial stability criteria are met if ${J_1^E-J_1^I>2}$ and ${J_1^E-\beta J_1^I<2}$, or, more simply, ${1<\frac{J_1^E-2}{J_1^I}<\beta}$. From these parameter sets we obtain the bifurcation diagram in figure \ref{fig:2Popmech}A. Similar as in the compressive nonlinearity (figure \ref{fig:1popCompNonLin}A and D) there is a finite regime of $C$ where the homogeneous state coexist with a bump state and for larger values of $C$ the bump state ceases to exist. As in the reduced model, this enables the switch-on to be performed with a tuned pulse of excitation and the switch-off to be done with a homogeneous excitatory transient input (figure \ref{fig:2Popmech}B); in this model (and in the other effective saturation models), however, none of the neurons reach their saturation rates. In the threshold difference model, for similar considerations, a bifurcation diagram as displayed in figure \ref{fig:2Popmech}C arises if ${0<\frac{J_1^E-2}{J_1^I}<1}$. Note that in the threshold difference model there also exists a regime where two homogeneous states co-exist with a bump state. Analogously to the reduced model, a bistability of homogeneous solutions occurs if the slope of the middle section of $F_0(I)$ is larger than 1; here, this slope is $J_0^E$. At the same time, since we require spatial instability of this solution, $J_1^E$ must be larger than 2. This immediately means that $J_0^E$ must also be larger than 2 (otherwise the connectivity function $J_E(\theta)$ is not strictly positive), hence in the  threshold difference model a bump state does not exist without bistability of homogeneous states. 

To conclude this part we demonstrate the effective saturation in a more general model, in which $\alpha_E<\alpha_I$, $\beta_E<\beta_I$ and $T_E<T_I$. Figure \ref{fig:2Popmech}D shows the bifurcation diagram of this model where, as in the previous examples, there exists a finite regime of $C$ where there is bistability of a homogeneous state and a bump, and for larger $C$ values the bump state vanishes.

\begin{figure}[!t]
\begin{center}
\includegraphics[width=6in]{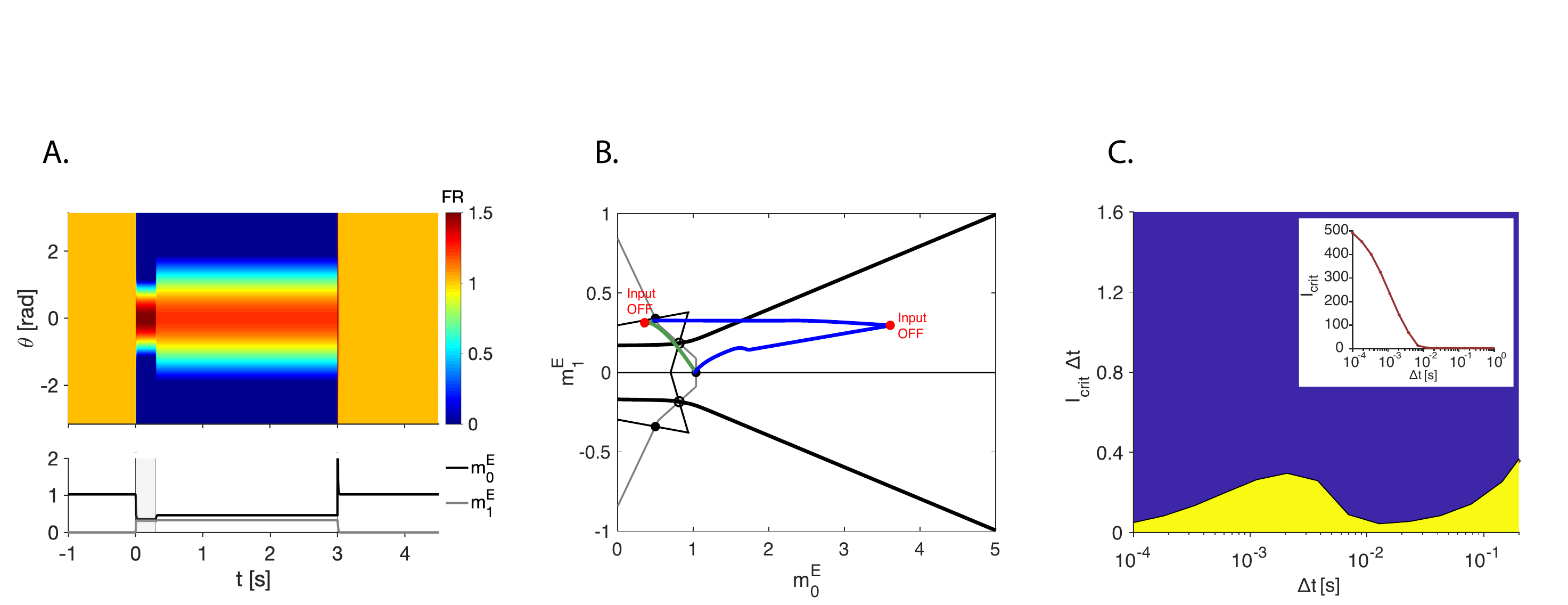}
\end{center}
\caption[Switching between states in the acceleration difference model]{
{\bf Switching between states in the acceleration difference model} Parameters as in figure \ref{fig:2Popmech}A. \textbf{A.} Dynamics of the network and order parameters in response to a 3 ms tuned input at $t = 0$ s centered at $\theta = 0$, and a 0.3 ms homogeneous input at $t = 3$ s. \textbf{B.} Dynamics of $m_0^E$ and $m_1^E$ from \textbf{A} plotted on the $m^E_0-m^E_1$ phase plane. Green: response to the switch-on input. Blue: response to switch-off input. \textbf{C.} Minimal switch-off input amplitude normalized by its duration, $I_{crit}\cdot\Delta t$, plotted against the duration of that input, $\Delta t$. The area in yellow depicts the region where the bump switch off cannot be performed. $\tau_E=5$ ms and $\tau_I=1$ ms.}
\centering
\label{fig:2popADshortDyn}
\end{figure}

In these models of effective saturation, for the same mechanism exposed previously, the switch-off can also be mediated by an input of duration shorter than the synaptic time constant. Figure \ref{fig:2popADshortDyn}A displays the response of the network to a 0.3 ms homogeneous input at t=3 s (same parameters as in figure \ref{fig:2Popmech}B). As showed, the bump can be switched off by this short stimulus. The increase of $m_0^E$ due to the excitatory external input pushes the activity trajectory described by $m_0^E$ and $m_1^E$ to cross the separatrix in the $m_0^E$ - $m_1^E$ phase space (figure \ref{fig:2popADshortDyn}B). Comparing the diagram of the minimal input current ($I_{crit}$) as a function of the input duration ($\Delta t$) for the compressive nonlinearity (figure \ref{fig:1popCompNonLinDyn}C) with the one from the two population acceleration difference model (figure \ref{fig:2popADshortDyn}C, inset), one can see that for long stimulus duration ($\Delta t>100$ ms) the $I_{crit}$ decays exponentially to a constant value in both scenarios. In the case of short stimulus duration ($\Delta t<10$ ms), for the compressive nonlinearity model there is an exponential increase of $I_{crit}$ as the stimulus duration is reduced while for the acceleration difference case the increase of $I_{crit}$ depicts a sigmoidal shape. In order to understand this behavior we represented the minimal input charge ($I_{crit}\cdot\Delta t$) as a function of the stimulus duration ($\Delta t$) (figure \ref{fig:2popADshortDyn}C). In this figure the boundary between the yellow and blue area depicts the minimal input charge needed to perform the switch-off. As the duration of the stimulus is reduced to a value of the same order of magnitude than the synaptic time constants ($\tau_E=5$ ms and $\tau_I=1$ ms) the minimal charge needed to do the switch-off increases to a maximum around $\Delta t=2$ ms. This behavior represents the effect of having two populations with different time constants. Since ${\tau_E>\tau_I}$, for stimulus duration ${\Delta t \approx\tau_E}$ the inhibitory firing rate increases faster than the excitatory. This is due to the fact that the external input is integrated faster for the inhibitory population and thus a larger input is needed to increase the excitatory firing rate (and thus $m_0^E$) in order to cross the separatrix. This effect is maximum for ${\tau_I<\Delta t<\tau_E}$. Once ${\Delta t \ll \tau_I}$ this effect is attenuated and the minimal charge tends to a saturation value for ${\Delta t}$ around ${ 0.1}$ ms.

\subsection*{Spiking network}

\begin{figure}[h!]
\begin{center}
\includegraphics[width=6.5in]{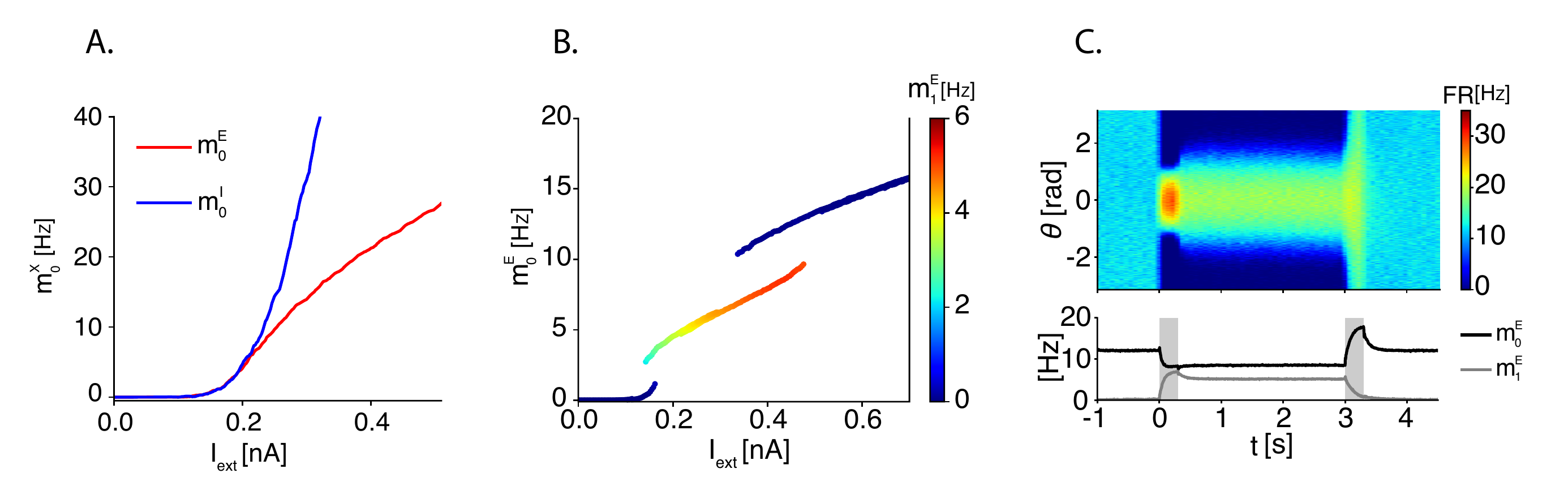}
\end{center}
\caption[Integrate and fire network model.]{
{\bf Integrate-and-fire network model.} \textbf{A.} Excitatory (red) and inhibitory (blue) neuronal transfer functions in the acceleration difference model. \textbf{B.} Bifurcation diagram of $m_0^E$ (y-axis) and $m_1^E$ (color code) with $I_{ext}$ as bifurcation parameter. Obtained by simulated annealing. \textbf{C.} Top: average rates of the neurons (color code) calculated in 75 ms time bins. Bottom: $m_0^E$ (black) and $m_1^E$ (gray). The network was subjected to a ${300\,\textrm{ms}}$ tuned input at $t = 0$ s a and a ${300\,\textrm{ms}}$ homogeneous excitatory input at $t = 3$ s. Parameters are specified in \cref{tab:IFPar,tab:IFParSyn}.
}
\centering
\label{fig:IntFire}
\end{figure}

We end with an implementation of the acceleration difference TF mechanism (figure \ref{fig:EffSat}A) in a network of integrate and fire neurons (see Methods section). We choose the parameters such that the inhibitory TF has a larger gain than the excitatory TF for input values above a certain threshold (figure \ref{fig:IntFire}A; parameters specified in table \ref {tab:IFPar}). The bifurcation diagram of the integrate and fire model (figure \ref{fig:IntFire}B) displays a bistable regime, where a homogeneous state co-exists with a bump state,  qualitatively similar to that described in the rate model. For large input values, only a stable homogeneous state exists, allowing the switch-off by global excitation. We show the switch-on and switch-off of the bump state in figure \ref{fig:IntFire}C, in which the firing rate of every neuron is calculated in 75 ms time bins. As in the rate model, a tuned input at ${t = 0\, \textrm{s}}$ causes the network to switch-on the bump state, and a 300 ms homogeneous excitatory input at ${t = 3\, \textrm{s}}$ switches the network back to a homogeneous state. Before the tuned input presentation, the excitatory and inhibitory neurons fire at rates of about 12 and 25 Hz, respectively. Close to the center of the bump during the delay period, the firing rates are around 20 Hz for the excitatory neurons and around 50 Hz for the inhibitory neurons (not shown here). We note that in this choice of parameters the excitatory synaptic decay time constant (see Methods) is larger than that of the inhibitory. Comparable excitatory and inhibitory synaptic time constants result in the homogeneous state losing stability. This arises from intrinsic neuronal dynamics, which are neglected in the rate model under the assumption that the membrane time constant is much smaller than the synaptic time constants. Indeed, when both excitatory and inhibitory synaptic time constants are enlarged respect to the membrane time constants, the bistability can be maintained for comparable excitatory and inhibitory synaptic time constants before the homogeneous state vanishes. 

\begin{figure}[H]
\begin{center}
\includegraphics[width=6in]{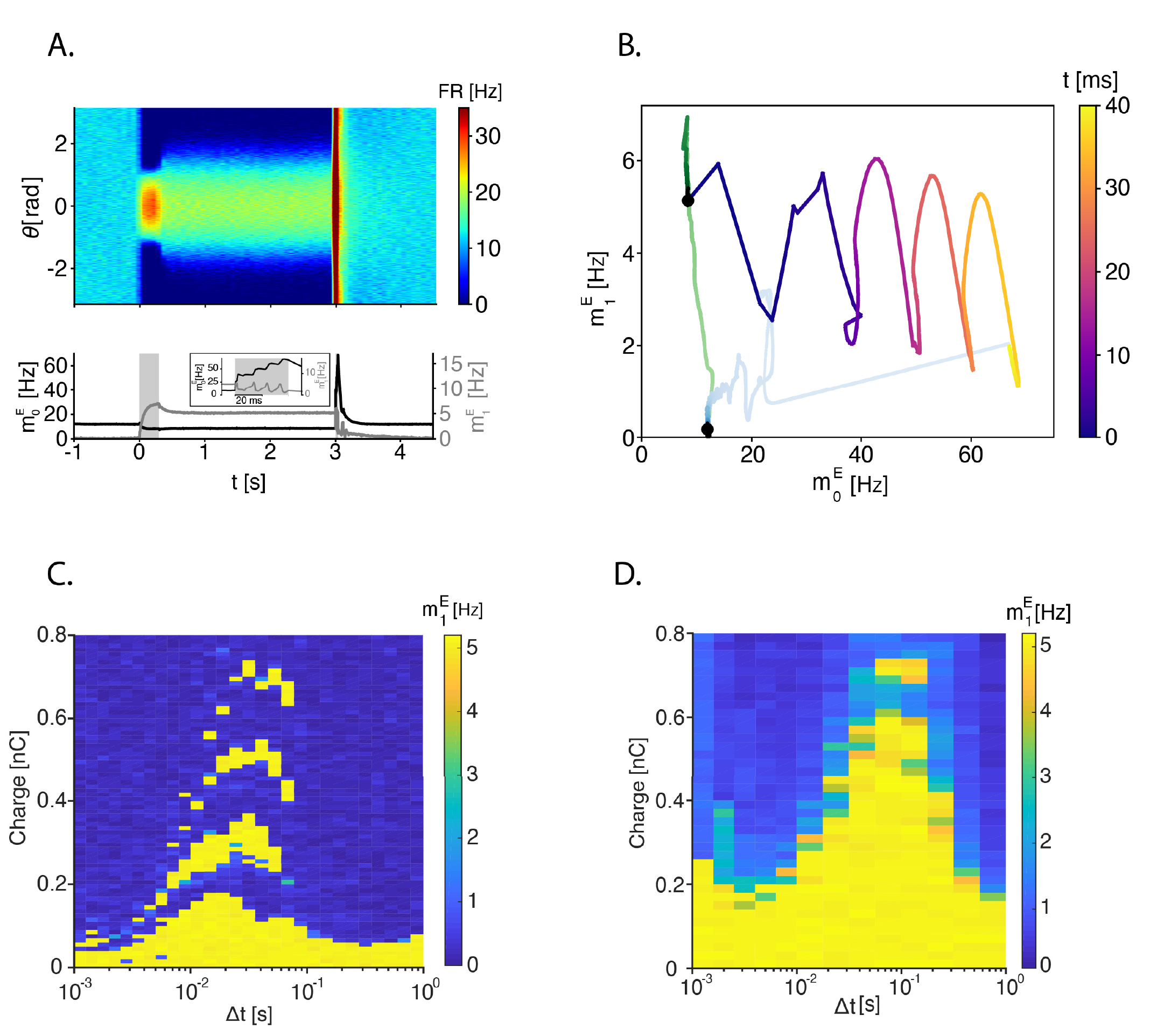}
\end{center}
\caption[Oscillatory transitions in the integrate and fire network model.]{
{\bf Oscillatory transitions in the integrate and fire network model.} \textbf{A.} Dynamics of the excitatory population of the LIF network subjected to a 300 ms tuned input at t=0 s and a 40 ms homogeneous input at t=3 s. Parameters as in figure \ref{fig:IntFire}. Top: average firing rate of the neurons calculated in 75 ms time bins. Bottom: Dynamics of the zero- and first order Fourier components of the average firing rate ($m_0^E$ and $m_1^E$, respectively). Inset: Detail of the response to the 40 ms homogeneous input at t=3 ms. \textbf{B.} Trajectory of $m_0^E$ and $m_1^E$ from A plotted on the  $m_0^E-m_1^E$ phase plane. Green: trajectory during switch on ($0<t<0.3$ s). Time course of the trajectory during 40 ms switch off color coded ($3<t<3.04$ s). Light Blue: trajectory during relaxation after the 40 ms stimulus. \textbf{C.} Stimulus input charge plotted against the duration of that input. Regions in blue represent where switch-off is performed and regions in yellow where the switch-off is not performed. \textbf{D.} Same as in \textbf{C} but with synaptic decay time constants increased by 6-fold.
}
\centering
\label{fig:IntFireTrans}
\end{figure}

The switch-off transition can be mediated by a short excitatory homogeneous input which restores the homogeneous state (figure \ref{fig:IntFireTrans}A), as predicted by the rate model. One striking difference of the spiking model regarding the rate model is that the dynamics of the order parameters $m_0^E$ and $m_1^E$ exhibits an oscillatory pattern during the transition (figure \ref{fig:IntFireTrans}A, inset). In figure \ref{fig:IntFireTrans}B we show that the trajectory of the activity in the $m_0^E-m_1^E$ phase plane during a 40-ms stimulus (color coded) depicts an oscillation due to alternating periods of bump state reactivation (large $m_1^E$) and a synchronous activation (small $m_1^E$). The mechanism underlying this oscillation is related to the membrane voltage dynamics. And it is exacerbated when the duration of the input and the synaptic time constants are of the order of the membrane time constants. Under the presentation of a homogeneous excitatory input of short duration, the membrane potential of the majority of the neurons (group 1 neurons) in the network reaches the firing threshold and thus, group 1 neurons spike simultaneously. Right after, the membrane potential of these neurons is restored to the reset potential. These group 1 neurons are set to an effective refractory period with duration $\tau_m \ln(\frac{(V_{th}-V_r)g_l}{I_{tot}})$. On the other hand, we define a group of neurons that had spiked just before the input arrived (group 2 neurons) and therefore were in the center of the bump. At the input onset, the group 2 neurons are in an effective refractory period and thus, they spike with a certain delay with respect to the input onset, creating a bump reactivation during the refractory period of group 1 neurons i.e. increasing $m_1^E$. 

The transition mechanism both in the rate and spiking model for short homogeneous input is based on the activity trajectory crossing the separatrix depicted in the $m_0^E-m_1^E$ phase plane. However, since in the spiking network model the activity trajectory is oscillatory, the crossing occurs several times and, depending on the value of $m_1^E$ at the time when the input is withdrawn the transition to the homogeneous state may or may not occur. For instance, if the input is withdrawn when the trajectory is above the separatrix (large $m_1^E$ i.e. bump reactivation), there will not be a transition to the homogeneous state. On the contrary, if the trajectory is below the separatrix at the moment of the input withdrawn (small $m_1^E$ i.e. synchronous activation) the trajectory will relax back to the homogeneous state. This behavior is well represented in the diagram depicting the minimal input charge needed to perform the switch-off as a function of stimulus duration in the spiking network (figure \ref{fig:IntFireTrans}C). The main differences, regarding the rate model, are the stratified regions (in yellow) for stimulus duration ${<0.1\,\textrm{s}}$ representing values of charge at which the switch-off cannot be performed. In the case in which the stimulus duration is much larger than the membrane and synaptic time constants (${\Delta t >0.1\,\textrm{s}}$ ), the amplitude of the oscillations is reduced due to the fading of the memory trace stored in the synaptic recurrent inputs. In this case the bump reactivation dissipates and the oscillation is damped. Therefore, the activity trajectory will cross the separatrix at most once and the maximum amplitude value of the oscillation will be always below the separatrix. Note that the yellow region at the bottom is qualitatively similar as in the rate model in figure \ref{fig:2popADshortDyn}C. And indeed, an increase of the synaptic time constants will retrieve a similar diagram in which oscillations no longer determine the transition (\ref{fig:IntFireTrans}D).

%
%
%
%
%
%
%
%
\section*{Discussion}
Working memory (WM) requires the representation of a past stimulus to persist over long times. In the present work we focus on visuospatial WM. We study, to a large extent analytically, how the interplay between neuronal nonlinearities and feature specific recurrent connectivity gives rise to neuronal dynamics during the delay period in this type of WM tasks.  In particular, we investigate the transient neuronal dynamics during switch-on and during switch-off of the persistent activity state.  

\subsection*{The two module model}

The simplest model we have considered consists of two modules. In one module the connectivity is nonspecific. Its activity can be persistent but it is always non-selective. In the second module the connectivity is functionally specific so that the activity is selective but not persistent. Correspondingly, each module in the system has a different role – one module holds the information about the stimulus and the other one enables the persistence of the representation. This functional organization resembles the relation that is believed to exist between the hippocampus and the cortex \cite{Eichenbaum1996}, where the representations are held in the cortex while the hippocampus supports the persistence. This scheme requires the existence of two types of neurons: one type codes for direction during delay periods, and the other is not direction-selective but codes in a binary way for whether or not a stimulus is stored in memory. Indeed, Funahashi et al. did report about \enquote{directional} and \enquote{omni-directional} neurons \cite{Funahashi1989}.

\subsection*{Selective persistent activity in a single recurrent network with functional specific connectivity}

Alternatively, selective persistent activity can emerge in a single recurrent network with specific connectivity. With sufficiently strong and spatially modulated recurrent excitation the network dynamics possess a continuous set (ring) of attractors. In these attractors the activity is \enquote{bumpy} and encodes for the cue direction. In order to account for the selective persistent activity observed experimentally, one must find the conditions under which a ring attractor can coexist with a homogeneous activity state. Such multistability requires appropriate nonlinearities. Using a simple modeling framework we investigated in a systematic manner, several mechanisms based on different f-I curve nonlinearities. We show that they differ by their bifurcation structures, the roles played by excitation and inhibition, as well as on the dynamics of the memory erasing after the delay period. 

\subsection*{Mechanisms underlying the switch-off of the persistent activity}

A simple and intuitive mechanism which can underlie the reset of the persistent activity at the end of the delay period, is a global inhibition of the excitatory neurons triggered by direct inhibition of the excitatory cells or by feedforward inhibition (e.g. \cite{Hansel2001,Wang1999,Roudi2007}). Physiologically, it is generally conceivable that in delayed-response tasks, either the response, the go signal or the reward triggers an increase in inhibition. In experiments, however, many neurons display a transient increase of their firing rate at the end of the delay period, even when the direction of the cue is away from the center of their tuning curve \cite{Funahashi1989,Constantinidis2001}. This suggests a global excitation of a large fraction of these neurons. Admittedly, these could be inhibitory neurons receiving excitation from outside the network, but nevertheless it is worthwhile to explore other possible switch off mechanisms.

In this work we describe two different mechanisms for switching-off the persistent state by global excitation. For stimulus duration much larger than the synaptic timescale, the switch-off relies on the bifurcation structure. Conversely, for stimulus duration shorter than the synaptic timescale, the switch off rely on the dynamics and specifically on the activity trajectory crossing of the separatrix. 

The mechanism majorly discussed here involves switch-off by a global prolonged excitatory pulse. This can only be implemented when the network response to input is saturating. In the case where the excitatory and inhibitory TFs have a similar shape, this requires the mechanism to make use of the saturating part of the TF. This means that during the excitatory pulse all the neurons are driven to the saturating part of the TF; there, the efficacy of recurrent excitation close to the center of the bump is no longer sufficient to fend off the inhibition arriving from the rest of the network, even though the activity around the center is higher than in the edges, and the bump is obliterated.
When the excitatory and inhibitory transfer functions differ substantially, this drop in relative efficacy of the excitation can emerge not because neurons arrive at the saturating part of the TF, but rather from a rise in the efficacy of the inhibition. At the risk of repetition, let us recall that in the model of Compte et al. \cite{Compte2000} the bump was also extinguished by a homogeneous excitatory pulse, and that saturating NMDA synapses in that model could potentially be the source of the increase in inhibition efficacy which causes the bump to disappear. A similar mechanism for memory erasure through global excitation is thoroughly described in a network dominated by inhibition in \cite{Brunel2001}.

The other discussed mechanism uses a brief, strong pulse of excitation delivered to the entire network (\cref{fig:1popExpNonLinDyn,fig:1popCompNonLinDyn,fig:2popADshortDyn,fig:IntFireTrans}) to extinguish the bump state. This pulse causes a transient increase in the average activity in the network, which is followed by a decay back to a homogeneous state. In terms of dynamics, the mechanism underlying this transition relies on the activity trajectory crossing the separatrix in the phase space. Notice that this mechanism allows the switch-off regardless of the nature of the nonlinearity in the TF and thus the bifurcation structure.
In the spiking network model studied in \cite{Gutkin2001,Laing2001}, a brief global excitation resets the activity back to its baseline level. This is because the brief pulse causes simultaneous firing in all the neurons. All the neurons enter then into a refractory period where they are unable to respond to inputs. Since the duration of this refractory period is larger than the time constant of the synapses (they use fast AMPA-like synapses), when the neurons are recovered from the refractory period, the synaptic memory trace has faded. Evidently, this is not the case in the model studied here, since the spike dynamics are not captured in the rate model and in the spiking model described here the timescale of the recurrent excitation is larger than the effective refractory period. Either way, these mechanisms inevitably raise the question of what could generate a systemic excitatory input that rises and decays with a time scale that is far smaller than that of the synapses in the network. A scheme of this sort could potentially be reliable in a network strongly dominated by slow NMDA synapses, although stability problems of over-excitation could arise. 

Another way to switch-off the selective persistent state by a short stimulus is to use an anti-tuned stimulus (figure \ref{fig:1popExpNonLinSwAntPhs}). In this case the input will create a bump of activity in the opposite direction making the activity trajectory to approach the lower half-plane in the phase space. If this input is withdrawn when the trajectory is somewhere inside the homogeneous BOA, the switch-off will succeed.

\subsection*{Interpretation of neuronal TF nonlinearities}

The behavior near firing threshold and close to the saturation of the neuronal TF are two key sources of nonlinearity. Around the threshold and in absence of noise, the activity of the neuron shows a sharp transition from quiescence to firing.  External noise smooths the transition resulting in an approximate power law behavior in some input range around the threshold \cite{Hansel2002,Miller2002,Persi2011}. The piecewise linear accelerating TFs we chose here (figure \ref{fig:ModDiag}B, top) mimics the response of a neuron in noisy environment which operates in this regime. Our simulations show that the mains conclusions of our analytical study also hold in leaky integrate-and-fire spiking network models in which the acceleration of the TF is smooth.

\subsection*{The interpretation of the rate model}

Our analytical investigations were performed in the rate-model framework (e.g. \cite{Ben-Yishai1995,Ermentrout1998b,Roudi2007,Roxin2005,Wilson1972}). This framework can be derived from first principles from the dynamics of a spiking network model if one assumes that the single neuron spiking dynamics is much faster than the synaptic dynamics.  In that case the firing rate of a neuron can be well approximated as an instantaneous function of its total input, namely, its f-I curve (see equation \ref{NeuOutput}). If one also assumes that the synaptic current induced by a presynaptic spike decay exponentially, the network follow the classical dynamics of a rate model in which the “rate” variables correspond to the synaptic inputs. 

A consequence of this interpretation is that the behaviors of the rate model and the spiking model, which tend to be in good correspondence when the synaptic time constants in the spiking model are large, show some discrepancies when the synapses are not slow enough. In the model presented here, although the general properties of spiking network behavior matched those in the rate-model network, one such inconsistency occurred: in the spiking model, for an inhibitory synaptic decay time constant of 5 ms, the stability of the bump broke down unless the excitation time constant was 4-5 times as large, whereas in the rate model, equal excitation and inhibition time constants were sufficient for bump stability. It should be noted though, that this discrepancy was made much smaller when both inhibition and excitation time constants were enlarged regarding the membrane time constants in the spiking network, i.e. approaching to the assumption taken in the rate model. This destabilization of the bump has also been reported in the spiking models considered by Compte et al. \cite{Compte2000}. In fact, in \cite{Wang1999,Compte2000} they show how slow and saturating NMDAR channels can stabilize a state of asynchronous persistent activity. Although this mechanism is supported by recent studies showing that persistent activity can be eliminated by blocking NMDAR channels \cite{Wang2013,VanVugt2020}, it is not clear whether this is a result of the reduction of the synaptic time scales or in the overall level of excitation. In fact, in a study by Hansel and Mato \cite{Hansel2001} is shown that slow excitation is not compulsory provided the inhibitory-to-inhibitory interactions are strong enough.

\subsection*{Effective saturation}

The mechanisms for persistence and selectivity in the model are based on the saturation of two effective functions of the input (equation \ref{Res2popIFP}). This saturation means that in the network, as the background level increases, the efficacy (or gain) of overall inhibition in the network increases as well (or, equivalently, the efficacy of excitation decreases). Note that it is not the level of inhibition that matters but the efficacy, that is, how strongly changes in the input affect the level of inhibition in the network. If the levels of excitation and inhibition are similarly affected by the input then any increase in inhibition would be accompanied by a respective increase in excitation, and the efficacy of inhibition will remain unchanged; to obtain the effective saturation, it is therefore necessary that the level of inhibition be increasingly more sensitive (relative to the excitation) to inputs. The present model uses nonlinearities in the TFs to obtain the effective saturation, although it is possible that a different form of effective saturation could also lead to similar mechanisms. An immediate candidate for a source of effective saturation is short term synaptic plasticity. The Compte et al. model \cite{Compte2000}, for instance, incorporates saturating NMDA synapses, which could induce an effective saturation that allows the emergence of multistability (as well as prolonged-excitation switch-off, as discussed above); still, it would likely be more efficient to study this hypothesis in a less detailed model that is easier to analyze.

\subsection*{Relation to balanced networks}

One final remark is with regard to the theory of balanced networks, first presented by van Vreeswijk and Sompolinsky in 1996 \cite{VanVreeswijk1996}. In this theory the neurons receive a bombardment of strong excitatory and inhibitory inputs, but a balance between them causes the total input to be close to the firing threshold of the neuron and the firing is driven by fluctuations in the input. A somewhat counter-intuitive result of that theory is that in the steady state of the balanced regime, population average firing rates obey a set of linear equations, regardless of the shape of the neuronal TFs. In other words, any nonlinearity in the TFs does not affect the macroscopic steady state of the network. Therefore, if one was to implement an attractor-based mechanism for WM in a balanced network, neuronal nonlinearities could not be of any assistance; alternatively, a compelling alternative source of nonlinearity for a balanced network is synaptic short-term plasticity \cite{Mongillo2012,Hansel2013}.

\newpage
\section*{Methods}
\subsection*{Rate model} \label{ssec:RateModel}
%
The model is comprised of $N_E$ excitatory neurons and $N_I$ inhibitory neurons. Each neuron receives a stimulus-dependent input $I_{stim}$ which is largest for some direction of the cue; this direction is termed the preferred direction (PD) of this neuron. The PDs are evenly distributed between $-\pi$ and $\pi$. In addition, each neuron receives a background input $C$ and a recurrent input $I_{rec}^X$ (where $X\in \{E,I\}$). The output of a neuron in population X with PD $\theta$ is its firing rate $r_X(\theta,t)$, which is computed by the neuron's input-output transfer function (TF), $g_X(I)$:
\begin{equation}\label{NeuOutput} 
r_X(\theta,t) = g_X\left(C+I_{rec}^X(\theta,t)+I_{stim}(\theta,t)\right) 
\end{equation}
In this work we study the effect of TF nonlinearities on network properties, namely the nonlinearities that stem from the threshold and saturation of the TF. We therefore consider piecewise-linear TFs of the form
\begin{equation} \label{GeneralTF}
g_X(I) = \left\{
\begin{array}{ll}
0,\\
\alpha_XI,\\ 
\beta_X(I-T_X)+\alpha_XT_X ,\\
\end{array}
\begin{array}{ll}
\quad I<0\\
\quad 0\leq I<T_X\\ 
\quad T_X\leq I\\
\end{array}
\right.
\end{equation}  
The variable $m_X(\theta,t)$ represents the activity in the outgoing synapses of the neuron in population $X$ with PD $\theta$. It has linear dynamics, with the neuron's firing rate as input:
\begin{equation}\label{mXt} 
\tau_X  \dot{m}_X(\theta,t)=-m_X(\theta,t)+r_X(\theta,t) 
\end{equation} 
The recurrent input is 
\begin{equation} \label{IrecX} 
I_{rec}^X(\theta,t) = \frac{1}{2\pi}\int_{2\pi}{J_{XE}(\theta-\xi)m_E(\xi,t)\mathrm{d}\xi}-\frac{1}{2\pi}\int_{2\pi}{J_{XI}(\theta-\xi)m_I(\xi,t)\mathrm{d}\xi}
\end{equation}
where $J_{XY}(\theta-\xi)$ is the weight of the connection from the neuron with PD $\xi$ in population $Y$ to the neuron with PD $\theta$ in population $X$. Throughout this paper we consider $J_{EY} = J_{IY} \triangleq J_Y$. We take the shapes of the connectivity functions to be 
\begin{equation}\label{JY} 
J_Y(\theta-\xi) = J^Y_0 + J^Y_1\mathrm{cos}(\theta-\xi)
\end{equation}
Considering this connectivity, the recurrent input in equation  (\ref{IrecX}) is independent of $X$ and can be written as:
\begin{equation}\label{Irec} 
I_{rec}(\theta,t) = J^E_0m^E_0(t)-J^I_0m^I_0(t)+\left(J^E_1m^E_1(t)-J^I_1m^I_1(t)\right)\mathrm{cos}(\theta)+\left(J^E_1m^E_2(t)-J^I_1m^I_2(t)\right)\mathrm{sin}(\theta)
\end{equation}
where
\begin{equation}\label{m0} 
m_0^Y(t) = \frac{1}{2\pi}\int_{2\pi}m_Y(\theta,t)\mathrm{d}\theta
\end{equation}
\begin{equation}\label{m1} 
m_1^Y(t) = \frac{1}{2\pi}\int_{2\pi}m_Y(\theta,t)\cos(\theta)\mathrm{d}\theta
\end{equation}
and
\begin{equation}\label{m2} 
m_2^Y(t) = \frac{1}{2\pi}\int_{2\pi}m_Y(\theta,t)\sin(\theta)\mathrm{d}\theta
\end{equation}
\subsubsection*{Steady states of the general case}
In the absence of $I_{stim}$, this network has two types of fixed points (FPs): a homogeneous FP, in which all the neurons fire at the same rate, and a \enquote{bump}, in which neurons within a certain vicinity fire at higher rates than all others. Owing to the symmetry of the network, we can address only steady state profiles which are symmetric around $\theta = 0$ without loss of generality. The general equations for the steady state are
\begin{equation}\label{SSgenTF} 
m_X(\theta) = g_X\left(C + I_0 + I_1\cos(\theta)\right)
\end{equation}
where $X\in\{E,I\}$ and $I_k \triangleq J_k^Em_k^E - J_k^Im_k^I$. We denote the total input to the neuron in the steady state by 
\begin{equation} \label{SSgenTF_Istar}
I^*(\theta) \triangleq C + I_0 + I_1\cos(\theta)
\end{equation}
In the case of the homogeneous steady state we have $I_1 = 0$, and the steady state equations become
\begin{equation}\label{SSgenTF_hom1} 
m_X(\theta) = g_X\left(C + I_0\right)
\end{equation}
Using the definition of $I_0$ we get that the steady state rate can be found by solving the equation
\begin{equation}\label{SSgenTF_hom2} 
I_0 = F_0(C+I_0)
\end{equation}
where $F_0(x)\triangleq J_0^Eg_E(x) - J_0^Ig_I(x)$ and then substituting $I_0$ in (\ref{SSgenTF_hom1}).

In the bump steady state, the solution can be characterized by three parameters: $\theta_0$, the maximal value of $\theta$ for which $I^*(\theta)$ is above zero; and $\theta_1^E$ and $\theta_1^I$, below which $I^*(\theta)$ is higher than $T_E$ and $T_I$, respectively. Substituting equation (\ref{SSgenTF}) in the definitions of the order parameters gives
\begin{equation}\label{SSgenTF_OP1} 
m_i^X(t) = (\beta_X-\alpha_X)\frac{1}{\pi}\int_{0}^{\theta_1^X}\left[I^*(\theta)-T_X\right]\cos(i\cdot \theta)\mathrm{d}\theta + \alpha_X \frac{1}{\pi}\int _{0}^{\theta_0} I^*(\theta)\cos (i \cdot \theta)\mathrm{d}\theta
\end{equation}
where $i\in\{0,1\}$. If $\theta_0$, $\theta_1^E$ and $\theta_1^I$ are all smaller than $\pi$ then the equations defining them are
\begin{equation}\label{SSgenTF_tt0} 
C + I_0 + I_1\cos(\theta_0) = 0
\end{equation}
and
\begin{equation}\label{SSgenTF_tt1X} 
C + I_0 + I_1\cos(\theta_1^X) = T_X
\end{equation}
From these equations we can obtain the relation
\begin{equation}\label{SSgenTF_eq1} 
\left(T_E - T_I\right)\cos(\theta_0) = T_E\cos(\theta_1^I) - T_I\cos(\theta_1^E) 
\end{equation}
Equation (\ref{SSgenTF_OP1}) can now be rewritten as 
\begin{equation}\label{SSgenTF_OP2} 
m_i^X = I_1\left((\beta_X-\alpha_X)f_i(\theta_1^X) + \alpha_Xf_i(\theta_0)\right)
\end{equation}
where
\begin{equation}\label{SSgenTF_fi} 
f_i(\theta) = \frac{1}{\pi}\int_0^\theta\left( \cos(\xi) - \cos(\theta)\right)\cos(i\cdot\xi)\mathrm{d}\xi
\end{equation}
From the definition of $I_1$ we can derive a second equation for $\theta_0$, $\theta_1^E$ and $\theta_1^I$:
\begin{equation}\label{SSgenTF_eq2} 
J_1^E(\beta_E - \alpha_E)f_1(\theta_1^E) - J_1^I(\beta_I - \alpha_I)f_1(\theta_1^I) + (J_1^E\alpha_E - J_1^I\alpha_I)f_1(\theta_0) = 1
\end{equation}
Taking equation (\ref{SSgenTF_OP2}) together with the definition of $I_0$ we can obtain:
\begin{equation}\label{SSgenTF_I0} 
I_0 = I_1\left[J_0^E(\beta_E - \alpha_E)f_0(\theta_1^E) - J_0^I(\beta_I - \alpha_I)f_0(\theta_1^I) + (J_0^E\alpha_E - J_0^I\alpha_I)f_0(\theta_0)\right]
\end{equation}
Substituting this in equations (\ref{SSgenTF_tt0}) and (\ref{SSgenTF_tt1X}) we can write a third equation for $\theta_0$, $\theta_1^E$ and $\theta_1^I$:
\begin{equation}\label{SSgenTF_eq3} 
J_0^E(\beta_E - \alpha_E)f_0(\theta_1^E) - J_0^I(\beta_I - \alpha_I)f_0(\theta_1^I) + (J_0^E\alpha_E - J_0^I\alpha_I)f_0(\theta_0) = -\frac{C}{T_E}\left[\cos(\theta_1^E)-\cos(\theta_0)\right] - \cos(\theta_0)
\end{equation}
After finding $\theta_0$, $\theta_1^E$ and $\theta_1^I$ from equations (\ref{SSgenTF_eq1}), (\ref{SSgenTF_eq2}) and (\ref{SSgenTF_eq3}) we can find the values of $I_k$ from equations (\ref{SSgenTF_tt0}) and (\ref{SSgenTF_tt1X}) and calculate the steady state rates from (\ref{SSgenTF}).

If $I^*(\theta)>0$ for all $\theta$ then equation (\ref{SSgenTF_tt0}) is no longer valid; instead, we substitute $\pi$ for $\theta_0$ in equation (\ref{SSgenTF_OP1}) to obtain
\begin{equation}\label{SSgenTF_OP0_BB} 
m_0^X = I_1(\beta_X-\alpha_X)f_0(\theta_1^X) + \alpha_X\left(C+I_0\right)
\end{equation}
and
\begin{equation}\label{SSgenTF_OP1_BB} 
m_1^X = I_1\left[(\beta_X-\alpha_X)f_1(\theta_1^X) + \frac{\alpha_X}{2}\right]
\end{equation}
From equation (\ref{SSgenTF_OP1_BB}) and the definition of $I_1$ we can get an equation for $\theta_1^E$ and $\theta_1^I$:
\begin{equation}\label{SSgenTF_BBeq1} 
J_1^E(\beta_E - \alpha_E)f_1(\theta_1^E) - J_1^I(\beta_I - \alpha_I)f_1(\theta_1^I) + \frac{1}{2}(J_1^E\alpha_E - J_1^I\alpha_I) = 1
\end{equation}
A second equation can be derived by combining equations (\ref{SSgenTF_OP0_BB}) and (\ref{SSgenTF_tt1X}) with the definition of $I_0$:
\begin{equation}\label{SSgenTF_BBeq2} 
\frac{J_0^E(\beta_E - \alpha_E)f_0(\theta_1^E) - J_0^I(\beta_I - \alpha_I)f_0(\theta_1^I)+\frac{C\left[\cos(\theta_1^E)-\cos(\theta_1^I)\right]}{T_E-T_I}}{1-(J_0^E\alpha_E - J_0^I\alpha_I)} = \frac{{T_I}\cos(\theta_1^E) - {T_E}\cos(\theta_1^I)}{T_E-T_I}
\end{equation}

By numerically solving these two equations we can find the steady state rates.

In a third case, $I^*(\theta)>T_E$ for all $\theta$ (here we assume that $T_E \leq T_I$). Now, substituting $\pi$ for $\theta_0$ and $\theta_1^E$ in (\ref{SSgenTF_OP1}) yields:
\begin{equation}\label{SSgenTF_SBB1} 
m_0^E = \beta_E(C+I_0-T_E)+\alpha_ET_E
\end{equation}
\begin{equation}\label{SSgenTF_SBB2} 
m_1^E = \beta_E I_1/2
\end{equation}
\begin{equation}\label{SSgenTF_SBB3} 
m_0^I = I_1(\beta_I-\alpha_I)f_0(\theta_1^I) + \alpha_I\left(C+I_0\right)
\end{equation}
and
\begin{equation}\label{SSgenTF_SBB4} 
m_1^I = I_1\left[(\beta_I-\alpha_I)f_1(\theta_1^I) + \frac{\alpha_I}{2}\right]
\end{equation}
Now using equations (\ref{SSgenTF_SBB2}) and (\ref{SSgenTF_SBB4}) and the definition of $I_1$ we can write an equation for $\theta_1^I$:
\begin{equation}\label{SSgenTF_BBeq1} 
J_1^E\beta_E/2-J_1^I\left[(\beta_I-\alpha_I)f_1(\theta_1^I)+\alpha_I/2\right]=1
\end{equation}
After numerically finding $\theta_1^I$ we can calculate $I_0$ and $I_1$ from equations (\ref{SSgenTF_SBB1}), (\ref{SSgenTF_SBB3}),  (\ref{SSgenTF_tt1X}) and the definition of $I_0$.
%

\subsubsection*{Reduction to one population} 
If the TFs (equation (\ref{GeneralTF})) and time constants are identical for excitatory and inhibitory neurons, then the dynamics (equation (\ref{mXt})) of both populations are the same. They can therefore be described by a single variable $m(\theta,t)$ with dynamics
\begin{equation}\label{mt} 
\tau  \dot{m}(\theta,t)=-m(\theta,t)+g\left(C+J_0m_0(t)+J_1m_1(t)\cos(\theta)+J_1m_2(t)\sin(\theta)+I_{stim}(\theta,t)\right) 
\end{equation} 
where $J_i = J^E_i - J^I_i$ and $m_i(t) = m_i^E(t) = m^I_i(t)$. This network was previously described in \cite{Ben-Yishai1995} with a threshold-linear TF; here, to investigate the effects of neuronal nonlinearities, we choose a TF as in equation (\ref{GeneralTF}) with $T_E = T_I = 1$, $\alpha_E = \alpha_I = 1$ and $\beta_E = \beta_I = \beta$: 
\begin{equation} \label{TF1pop}
g(I) = \left\{
\begin{array}{ll}
0,\\
I,\\ 
\beta(I-1)+1 ,\\
\end{array}
\begin{array}{ll}
\quad I<0\\
\quad 0\leq I<1\\ 
\quad 1\leq I\\
\end{array}
\right.
\end{equation}  
Here too we address only steady states profiles that are symmetric around $\theta = 0$; the general FP equation is therefore:
\begin{equation}\label{BumpFP} 
m(\theta)=g\left[C+J_0m_0+J_1m_1\cos(\theta)\right] 
\end{equation} 
In the homogeneous state we have $m_1 = 0$, and the rate is determined by solving the equation 
\begin{equation}\label{HomFP} 
I^*-C = J_0g(I^*)
\end{equation} 
where $I^* = C + J_0m_0$. As in \cite{Ben-Yishai1995}, the condition for the stability of this FP is that $J_0g'(I^*)<1$ and $J_1g'(I^*)<2$ (where $g'(I^*)$ is the derivative of $g(I)$ at the point $I^*$). In the results section we analyze the difference between the cases $\beta>1$ and $\beta<1$ (figures \ref{fig:1popExpNonLin} and \ref{fig:1popCompNonLin}).
\newline
 In the reduced model, the bump solution is characterized by only two angles, since $T_E = T_I$ and therefore $\theta_1^E = \theta_1^I \triangleq \theta_1$. If we define $G_i(\theta_0,\theta_1) = \frac{1}{\pi}\left(\int_0^{\theta_0}{\cos^i(\theta)\mathrm{d}\theta}+(\beta-1)\int_0^{\theta_1}{\cos^i(\theta)\mathrm{d}\theta}\right)$, and a matrix
\begin{equation} \label{A} 
\mathbf{A}\left(\theta_0, \theta_1\right) = 
\left[\begin{array}{ll}
G_0\left(\theta_0,\theta_1\right)J_0-1 &  G_1\left(\theta_0,\theta_1\right)J_1\\
G_1\left(\theta_0,\theta_1\right)J_0 &  G_2\left(\theta_0,\theta_1\right)J_1-1\\
\end{array}\right]
\end{equation} 
then $m_0$ and $m_1$ obey the linear relation:
\begin{equation}\label{BB2}
{\mathbf {A}}\left(\theta_0, \theta_1\right) 
\cdot
\left[\begin{array}{c}
m_0\\
m_1
\end{array}\right]
=
\frac{(\beta-1)}{\pi}\left[\begin{array}{c}
\theta_1\\
\sin(\theta_1)\end{array}\right] -
C\cdot\left[\begin{array}{ll}
G_0\left(\theta_0,\theta_1\right)\\
G_1\left(\theta_0,\theta_1\right) 
\end{array}\right]
\end{equation} 
In the case where $\theta_0$ and $\theta_1$ are smaller than $\pi$, they can be found by numerically solving the equations 
\begin{equation}\label{BumpFP_tt1} 
J_0\left(f_0(\theta_0)+(\beta-1)f_0(\theta_1)\right) = C\cdot\left( \cos(\theta_0)-\cos(\theta_1)\right)-\cos(\theta_0)
\end{equation} 
and 
\begin{equation}\label{BumpFP_tt2} 
J_1\left(f_1(\theta_0)+(\beta-1)f_1(\theta_1)\right) =1
\end{equation} 
where $f_i(\theta)$ are as defined in equation (\ref{SSgenTF_fi}). In the case where the bump is above zero for all $\theta$, the value of $\theta_1$ can be found by substituting $\pi$ in $\theta_0$ in equation (\ref{BumpFP_tt2}) and solving numerically for $\theta_1$. 

Not for every $C$ value there exists a bump solution. To find the critical values of $C$ we use equation (\ref{BumpFP_tt1}) to write $C$ as a function of $\theta_0$ and $\theta_1$, and then search for extrema of this function under the constraint given by equation (\ref{BumpFP_tt2}). We find that the values of $\theta_i$ at the extrema can be found by numerically solving equation (\ref{BumpFP_tt2}) with the equation
\begin{equation}\label{Cmin1} 
\Psi_0+\left(\beta-1\right)\Psi_1 = 0
\end{equation} 
where
\begin{equation}\label{Cmin2} 
\Psi_0 = \left[J_0\left(f_0(\theta_0) + (\beta-1)\frac{\sin(\theta_1)-\theta_1\cos(\theta_0)}{\pi}\right)+\cos(\theta_0)\right]\cdot\sin(\theta_0)
\end{equation} 
and
\begin{equation}\label{Cmin3} 
\Psi_1 = \left[J_0\left(\frac{\sin(\theta_0)-\theta_0\cos(\theta_1)}{\pi}+ (\beta-1)f_0(\theta_1) \right)+\cos(\theta_1)\right]\cdot\sin(\theta_1)
\end{equation} 

After finding $\theta_i$ we can substitute them in equation (\ref{BumpFP_tt1}) to find the critical value of $C$.

\bigskip

In another case where the bump does not exist we have $m_i \rightarrow \infty$. The critical value for this case is obtained by taking the limit $\theta_1 \rightarrow \theta_0$. Substituting $\theta_0 = \theta_1 \triangleq \theta_c$ in equation (\ref{BumpFP_tt1}) we obtain the equation
\begin{equation}\label{theta_c} 
\beta J_0\frac{1}{\pi}\left(\theta_c-\tan(\theta_c)\right) = 1
\end{equation} 
from which we can numerically find the value of $\theta_c$. Substituting $\theta_0$ and $\theta_1$ for $\theta_c$ in equation (\ref{BumpFP_tt2}) will then give us the maximal value of $J_1$ for which the bump exists; above it the system diverges (note that this value depends only on $J_0$ and not on $C$).

To analyze the stability of these steady states we need to consider the response of the network to a small deviation from the FP, such that $m\left(\theta,t\right) = m^*\left(\theta\right)+\delta m\left(\theta,t\right)$, where $m^*(\theta)$ satisfies equation (\ref{BumpFP}). In this case, the dynamics is governed by: 
\begin{equation} \label{dm_dot}  \tau\dot{\delta m}(\theta,t)= -\delta m(\theta,t)+ g'\left(I^*(\theta)\right)\cdot  \left[J_0\delta m_0(t)+J_1\delta m_1(t)\cos(\theta)+J_1\delta m_2(t)\sin(\theta)\right] \end{equation} 
where 
\begin{equation} \label{dm0} 
\delta m_0(t) = \frac{1}{2\pi}\int_{2\pi}\delta m(\theta,t)\mathrm{d}\theta 
\end{equation}
\begin{equation} \label{dm1} 
\delta m_1(t) = \frac{1}{2\pi}\int_{2\pi}\delta m(\theta,t)\cos(\theta)\mathrm{d}\theta
\end{equation}
\begin{equation} \label{dm2} 
\delta m_2(t) = \frac{1}{2\pi}\int_{2\pi}\delta m(\theta,t)\sin(\theta)\mathrm{d}\theta
\end{equation}
$I^*$ is the total input at the FP, and $g'(I)$ is the derivative of $g(I)$: 
\begin{equation} \label{gg}
g'(I) = \left\{
\begin{array}{ll}
0,\\
1,\\ 
\beta,\\
\end{array}
\begin{array}{ll}
\quad I<0\\
\quad 0\leq I\leq1\\ 
\quad 1<I\\
\end{array}
\right.
\end{equation}
Using equation (\ref{dm_dot}) in differentiating equations (\ref{dm0}) and (\ref{dm1}) with respect to time we obtain the dynamics of $\delta m_0$ and $\delta m_1$: 
\begin{equation}\label{dm01_dot}
\tau \frac{\mathrm{d}}{\mathrm{d}t}
\left[\begin{array}{ll}
\delta m_0\\
\delta m_1
\end{array}\right] = 
\mathbf{A}\left(\theta_0, \theta_1\right)
\cdot 
\left[\begin{array}{ll}
\delta m_0\\
\delta m_1
\end{array}\right]
\end{equation}
where $\mathbf{A}$ is as in equation (\ref{A}). Similarly, and considering equation (\ref{BumpFP_tt2}), we find that $\dot{\delta m_2} = 0$; this is not surprising, since the symmetry of the network implies that the phase of the bump at the steady state is arbitrary. 
The only requirement for stability is therefore that the eigenvalues of the matrix $\mathbf{A}(\theta_0,\theta_1)$
at the steady state $m(\theta)$ must be negative. This is fulfilled if:
\begin{equation} \label{StabCond1}
 G_0\left(\theta_0, \theta_1\right)\cdot \left(J_0+J_1\right) < 3
\end{equation}
and
\begin{equation} \label{StabCond2}
 \left(J_0G_0\left(\theta_0, \theta_1\right)-1\right) \cdot \left(J_1G_2\left(\theta_0, \theta_1\right)-1\right) - 
J_0J_1G_1\left(\theta_0, \theta_1\right)^2 > 0
\end{equation}

To plot the flow diagram in figures \ref{fig:1popExpNonLinDyn}B, \ref{fig:1popExpNonLinSwAntPhs}B and \ref{fig:1popCompNonLinDyn}B we write the dynamics of $m_k$ in the absence of an external stimulus by Fourier transforming equation (\ref{mt}):
\begin{equation}\label{m0t} 
\tau  \dot{m_0}=-m_0+\frac{1}{2\pi}\int_{2\pi}{g\left(C+J_0m_0+J_1m_1\cos(\theta)+J_1m_2\sin(\theta)\right)\mathrm{d}\theta} 
\end{equation} 
\begin{equation}\label{m1t} 
\tau  \dot{m_1}=-m_1+\frac{1}{2\pi}\int_{2\pi}{g\left(C+J_0m_0+J_1m_1\cos(\theta)+J_1m_2\sin(\theta)\right)\cos(\theta)\mathrm{d}\theta} 
\end{equation} 
\begin{equation}\label{m2t} 
\tau  \dot{m_2}=-m_2+\frac{1}{2\pi}\int_{2\pi}{g\left(C+J_0m_0+J_1m_1\cos(\theta)+J_1m_2\sin(\theta)\right)\sin(\theta)\mathrm{d}\theta} 
\end{equation} 
The flow on the $m_0-m_1$ plane reflects the case where the profile is completely symmetric around $\theta = 0$, hence $m_2 = 0$; moreover, from the symmetry of the system we can tell that the flow is symmetric about the $m_0$ axis and therefore the lower half plane is a mirror image of the upper half plane. In the upper half plane the flow is given by  
\begin{equation}\label{m01_dot}
\tau \frac{\mathrm{d}}{\mathrm{d}t}
\left[\begin{array}{ll}
m_0\\
m_1
\end{array}\right] = 
{\mathbf {A}\left(\theta_0, \theta_1\right) }
\cdot
\left[\begin{array}{c}
m_0\\
m_1
\end{array}\right]
-
\frac{(\beta-1)}{\pi}\left[\begin{array}{c}
\theta_1\\
\sin(\theta_1)\end{array}\right] 
+
C\cdot\left[\begin{array}{ll}
G_0\left(\theta_0,\theta_1\right)\\
G_1\left(\theta_0,\theta_1\right) 
\end{array}\right]
\end{equation}
where $\theta_k$ is computed by defining $u_k = \frac{k-C-J_0m_0}{J_1m_1}$ and then
\begin{equation} \label{tt_k}
\theta_k = \left\{
\begin{array}{ll}
\pi,\\
\cos^{-1}(u_k),\\ 
0,\\
\end{array}
\begin{array}{ll}
\quad u_k<-1\\
\quad -1\leq u_k\leq1\\ 
\quad 1<u_k\\
\end{array}
\right.
\end{equation}
To find the stable branch of the unstable FP we first numerically calculate the normalized eigenvectors of $\mathbf{A}$ at this FP. Then, we use equation (\ref{m01_dot}) to track the trajectory that starts from this FP in the stable eigendirection (the direction of the eigenvector with the negative eigenvalue) backwards in time. 
%

\subsubsection*{Homogeneous state stability in the general model} 
As in the reduced model, we examine the dynamics of the network following a small perturbation from the homogeneous FP such that $m_X(\theta,t) = m_X^* + \delta m_X(\theta,t)$ (where $m_X^*$ is the homogeneous steady state rate of population $X$, $X\in \{E,I\}$). The dynamics then follows 
\begin{equation} \label{dmX_dot}  
\tau_X\dot{\delta m_X}(\theta,t)= -\delta m_X(\theta,t)+  g_X'\left(I^*\right)\cdot \delta I(\theta,t) 
\end{equation} 
where
\begin{equation} \label{dI}  \delta I(\theta,t) = J^E_0\delta m^E_0(t)-J^I_0\delta m^I_0(t)+\left(J^E_1\delta m^E_1(t)-J^I_1\delta m^I_1(t)\right)\cos(\theta)+\left(J^E_1\delta m^E_2(t)-J^I_1\delta m^I_2(t)\right)\sin(\theta)
\end{equation} 
the derivative of $g_X(I)$ is
\begin{equation} \label{dgX}
g_X'(I) = \left\{
\begin{array}{ll}
0,\\
\alpha_X,\\ 
\beta_X,\\
\end{array}
\begin{array}{ll}
\quad I<0\\
\quad 0\leq I<T_X\\ 
\quad T_X\leq I\\
\end{array}
\right.
\end{equation}  
and the order parameters $\delta m_i^X$ are
\begin{equation} \label{dm0_X} 
\delta m_0^X(t) = \frac{1}{2\pi}\int_{2\pi}\delta m_X(\theta,t)\mathrm{d}\theta 
\end{equation}
\begin{equation} \label{dm1_X} 
\delta m_1^X(t) = \frac{1}{2\pi}\int_{2\pi}\delta m_X(\theta,t)\cos(\theta)\mathrm{d}\theta
\end{equation}
\begin{equation} \label{dm2_X} 
\delta m_2^X(t) = \frac{1}{2\pi}\int_{2\pi}\delta m_X(\theta,t)\sin(\theta)\mathrm{d}\theta
\end{equation}
Let us define $\widetilde{J}_i^{XY}$ to be $g_X'(I^*)J_0^Y$ if $i = 0$ and $ g_X'(I^*)J_1^Y/2$ otherwise. Then, substituting the order parameters in equation (\ref{dmX_dot}) we obtain
\begin{equation} \label{dmX_vector} 
\frac{\mathrm{d}}{\mathrm{d}t}
\left[\begin{array}{c}
\delta m^E_i\\
\delta m^I_i
\end{array}\right]= 
\left[\begin{array}{ll}
\left( \widetilde{J}_i^{EE}-1\right)/\tau_E 	&    {-}\widetilde{J}_i^{EI}/\tau_E 		\\
\widetilde{J}_i^{IE}/\tau_I 	&   {-\left( \widetilde{J}_i^{II}+1\right)}/\tau_I 		\\
\end{array}\right]\cdot
\left[\begin{array}{c}
\delta m^E_i\\
\delta m^I_i
\end{array}\right]
\end{equation} 
The stability condition is that the eigenvalues of the matrix must be negative; for this, the determinant has to be positive and the trace must be negative. The conditions for the determinant are:  
\begin{equation} \label{StabCondGen1} 
g_E'(I^*)J_0^E- g_I'(I^*)J_0^I<1
\end{equation}
 and 
\begin{equation} \label{StabCondGen2} 
 g_E'(I^*)J_1^E-g_I'(I^*)J_1^I<2
\end{equation}
These conditions are independent of the time constants $\tau_E$ and $\tau_I$. The conditions for the trace to be negative are 
\begin{equation} \label{StabCondGen3} 
\tau_E/\tau_I>\frac{g_E'(I^*)J_0^E-1}{ g_I'(I^*)J_0^I+1}
\end{equation}
and
\begin{equation} \label{StabCondGen4} 
\tau_E/\tau_I>\frac{g_E'(I^*)J_1^E/2-1}{ g_I'(I^*)J_1^I/2+1}
\end{equation}
In the case $\tau_E = \tau_I$, if the conditions for the determinant are fulfilled then so are the conditions for the trace.

\subsubsection*{Bump state stability in the general model} 
Here too we refer to equation (\ref{dmX_dot}), but with $I^*$ as given in equation (\ref{SSgenTF_Istar}). We define the steady state profile values:
\begin{equation} \label{StabGenTF_KiX}
K_i^X = \frac{1}{\pi}\left(\alpha_X\int_0^{\theta_0}{\cos^i(\theta)\mathrm{d}\theta}+(\beta_X-\alpha_X)\int_0^{\theta_1^X}{\cos^i(\theta)\mathrm{d}\theta}\right)
\end{equation}
Then, using equation (\ref{dmX_dot}) and the definitions of $\delta m_0^X$ and $\delta m_1^X$ we can derive the linearized dynamics
\begin{equation} \label{StabGenTF_LinDyn}
\tau_E \frac{d}{dt}\left[\begin{array}{c}
\delta m^E_0 	\\
\delta m^I_0 	\\
\delta m^E_1 	\\
\delta m^I_1
\end{array}\right]= 
\mathbf{K} \cdot \left[\begin{array}{c}
\delta m^E_0 	\\
\delta m^I_0 	\\
\delta m^E_1 	\\
\delta m^I_1
\end{array}\right]
\end{equation} 
where
\begin{equation} \label{StabGenTF_K}
\mathbf{K} \triangleq 
\left[\begin{array}{llll}
J_0^EK_0^E-1				&	-J_0^IK_0^E 					& 	J_1^EK_1^E 					&	 	-J_1^IK_1^E 				\\
\frac{\tau_E}{\tau_I}J_0^EK_0^I 	&	-\frac{\tau_E}{\tau_I}(J_0^IK_0^I+1) 	& 	\frac{\tau_E}{\tau_I}J_1^EK_1^I 		&		-\frac{\tau_E}{\tau_I}J_1^IK_1^I 	\\
J_0^EK_1^E 				&	-J_0^IK_1^E 					& 	J_1^EK_2^E-1 					&	 	-J_1^IK_2^E 				\\
\frac{\tau_E}{\tau_I}J_0^EK_1^I 	&	-\frac{\tau_E}{\tau_I}J_0^IK_1^I 	 	& 	\frac{\tau_E}{\tau_I}J_1^EK_2^I 		&		-\frac{\tau_E}{\tau_I}(J_1^IK_2^I+1) \\
\end{array}\right]
\end{equation}
The bump state is only stable if all the eigenvalues of $\mathbf{K}$ have negative real parts. These eigenvalues can be found numerically. The definition of $\delta m_2^X$ together with equation  (\ref{dmX_dot}) yields linearized dynamics for $\delta m_2^X$:
\begin{equation} \label{StabGenTF_LinDyn_sin}
\tau_E\frac{d}{dt}\left[\begin{array}{c}
\delta m^E_2 	\\
\delta m^I_2 	\\
\end{array}\right]= \mathbf{B}
\cdot \left[\begin{array}{c}
\delta m^E_2 	\\
\delta m^I_2 	\\
\end{array}\right]
\end{equation} 
where
\begin{equation}\label{StabGenTF_B}
\mathbf{B}\triangleq
\left[\begin{array}{ll}
J_1^E\left[(\beta_E-\alpha_E)f_1(\theta_1^E)+\alpha_Ef_1(\theta_0)\right]-1				&	-J_1^I\left[(\beta_E-\alpha_E)f_1(\theta_1^E)+\alpha_Ef_1(\theta_0)\right] 					\\
\frac{\tau_E}{\tau_I}J_1^E\left[(\beta_I-\alpha_I)f_1(\theta_1^I)+\alpha_If_1(\theta_0)\right] 	&	-\frac{\tau_E}{\tau_I}\left(J_1^I\left[(\beta_I-\alpha_I)f_1(\theta_1^I)+\alpha_If_1(\theta_0)\right]-1\right)  		\\
\end{array}\right]
\end{equation}
In order for the eigenvalues of $\mathbf{B}$ to be negative, the determinant of $\mathbf{B}$ must be non-negative and the trace must be negative. In this case, because of the steady state equation (\ref{SSgenTF_eq2}) we find that the determinant is always zero. The demand for the negative trace leads to the condition
\begin{equation}\label{StabGenTF_PhasCond}
\tau_E/\tau_I>\frac{J_1^E\left[(\beta_E-\alpha_E)f_1(\theta_1^E)+\alpha_Ef_1(\theta_0)\right]-1}{J_1^I\left[(\beta_I-\alpha_I)f_1(\theta_1^I)+\alpha_If_1(\theta_0)\right]+1}
\end{equation}
%

\bigskip

\subsection*{Integrate and fire model}\label{sec:IFModel}
We consider a network of $N_E$ excitatory and $N_I$ inhibitory integrate-and-fire neurons, connected in an all-to-all manner.  Neurons, as in the rate model, are labeled according to their PDs, and the variable $V_X(\theta,t)$ represents the membrane potential of the neuron with PD $\theta$ in population $X$ ($X\in\{E,I\}$). The evolution of $V_X(\theta,t)$ in the subthreshold regime follows:
\begin{equation}\label{IF_Vdot} 
C_X\dot{V}_X(\theta,t)=-g_l(V_X(\theta,t)-v_l)+I_{rec}^E(\theta,t)-I_{rec}^I(\theta,t)+I_{bg}(\theta,t)+I_{stim}(\theta,t) 
\end{equation} 
where $C_X$ is the membrane capacitance, $g_l$ and $v_l$ are the leak conductance and reversal potential, respectively, $I_{rec}^X$ is the recurrent input from population $X$, $I_{bg}$ is a background input and $I_{stim}$ is a stimulus dependent input. Whenever $V_X$ reaches a threshold $v_t^X$ a spike is emitted and the membrane potential is reset to $v_r^X$, without a refractory period. The leak conductance $g_l$ and the leak reverse potential $v_l$ were identical in both populations. The parameter set we used is detailed in table \ref{tab:IFPar}.
%
%
%
%
%
%
%
\subsubsection*{Intrinsic neuron properties }

Membrane parameters are similar as in \cite{McCormick1985}, additionally we make sure all the values we use lie on the biological range by checking electrophysiology databases such as NeuroElectro.org \cite{NeuroElectro}. Reset potential after the spike for the excitatory is often taken to be more negative ($\approx$ -90 mV) than the inhibitory ($\approx$ -60 mV). Since the excitatory membrane time constant ($\tau_E$=20 ms) is typically twice the inhibitory membrane time constant ($\tau_I$=10 ms), the more negative reset potential will introduce an effective refractory period accounting for the fact that excitatory neurons usually display spike-frequency adaptation \cite{Connors1990}.

\subsubsection*{Recurrent inputs} 
When a spike occurs at time $t_{sp}$ in a presynaptic neuron with PD $\varphi$ in population $Y$, the current evoked in the postsynaptic neuron with PD $\theta$ in population $X$, also known as postsynaptic current (PSC) is:

\begin{equation}
I_{sp}^{Y}(\theta-\varphi,t-t_{sp})=J_{Y}(\theta-\varphi)\, s_{Y}(t-t_{sp})
\label{isp}
\end{equation}

where $J_{Y}(\theta-\varphi)$ is the total charge transferred in the synapse due to a single presynaptic spike, and it represents the synaptic strength (units of charge). This strength is scaled inversely with the size of the presynaptic population. As in the rate model, the connectivity is taken to have a cosine shape and to be independent on the nature of the postsynaptic population $X$ (i.e. ${J_{EE}=J_{EI}\triangleq J_E}$ and ${J_{II}=J_{IE}\triangleq J_I}$):

\begin{equation}
J_Y(\theta-\varphi)=\frac{1}{N_Y}\,(J_0^Y+J_1^Y\, \cos(\theta-\varphi))
\label{jy}
\end{equation}

On the other hand, the quantity $s_Y(t-t_{sp})$ describes the time course of the PSC in the postsynaptic neuron elicited by a spike at time $t_{sp}$ from a presynaptic neuron in population $Y$. It has units of $[1/s]$ and it is described by a dual exponential waveform describing the fast dynamics at the opening and the slow dynamics at the closing of the synaptic receptors \cite{Sterratt2011}:

\begin{equation}
s_Y(t-t_{sp})=\frac{1}{\tau_d^Y-\tau_r^Y} \left [ \exp \left ( {\frac{-(t-t_{sp})}{\tau_d^Y}} \right )- \exp \left ( {\frac{-(t-t_{sp})}{\tau_r^Y}} \right ) \right ]
\label{IF_s}
\end{equation}

Where $\tau_r^Y$ and $\tau_d^Y$ are the synaptic rise and decay time constants, respectively. The normalizing prefactor ensures that the total area under the PSC curve $s_Y$ generated by a single spike is equal to 1. The recurrent input from the presynaptic population $Y$ to the neuron with PD $\theta$ in the postsynaptic population $X$ is defined as the sum over all the currents evoked by all the presynaptic neurons:

\begin{equation}
I_{rec}^{YX}(\theta,t)=\sum_i\sum_jI_{sp}^Y (\theta-\varphi_i,t-t_j(\varphi_i))
\label{irec}
\end{equation}

where $t_j(\varphi_i)$ is the time of the j-th spike fired by the presynaptic neuron with PD $\varphi_i$.

\subsubsection*{Synaptic currents}

For a more biologically realistic implementation of the synaptic currents we consider that excitatory synaptic inputs are a combination of fast AMPA and slow NMDA currents (\cref{tab:IFParSyn}). By using biological parameters for the rise and decay time constants in equation \ref{IF_s} we describe the time course of the AMPA and NMDA EPSC separately. These dynamics represents the ion flow occurring in the synapse during the opening and closing of AMPAR and NMDAR. For the AMPA EPSC we chose a rise time constant ($\tau_r^{AMPA}$) of 0.5 ms and a decay time constant ($\tau_d^{AMPA}$) of 5 ms \cite{Kleppe1999, Gonzalez-Burgos2008}. For the rise and decay NMDA EPSC we chose $\tau_r$=1 ms and $\tau_d$=50 ms, respectively \cite{Jonas1993, Gonzalez-Burgos2008}. We found in the literature that in the excitatory to excitatory synapse ($E-E$) the mean NMDA/AMPA ratio calculated using the area under the EPSC in dlPFC neurons in monkey is 2.7 \cite{Gonzalez-Burgos2008}. On the other hand, the NMDA/AMPA ratio in the excitatory to inhibitory synapse ($E-I$) is 0.5 \cite{Wang-Gao2009}. This value is computed by taking the NMDA and AMPA charge (EPSC area) of the fast spiking interneurons and taking into account that, in the adult rat, only the 26\% of this interneurons present NMDA currents. To calculate the resultant synaptic current of the contribution of NMDA and AMPA currents we proceed as follows:		
\begin{equation}\label{IF_Irec_NMDA} 
I_{rec}^{EX}(\theta,t)=\frac{r^{EX}}{r^{EX}+1}\cdot I_{rec}^{NMDA}(\theta,t)+\frac{1}{r^{EX}+1}\cdot I_{rec}^{AMPA}(\theta,t)
\end{equation}

Where $r^{EX}$ is the NMDA/AMPA ratio for $E-X$ the synapses.

\subsubsection*{Background input} 
Each neuron receives a noisy background input of the form:
\begin{equation}\label{IF_Ibg} 
I_{bg}(\theta,t) = \mu_{bg} + \sigma_{bg}\cdot\xi(t)
\end{equation} 
where $\xi(t)$ is a Gaussian white noise with zero correlation in time and between neurons. Since our interest is in the effect of the difference between excitatory and inhibitory neurons' TFs on network states, we add a quantity $\Delta_{bg}$ to the background input of the excitatory neurons, in order to effectively induce a difference in the threshold of excitatory and inhibitory TFs.   

\newpage
\begin{table}[H]%
\centering %
\begin{tabular}{@{} c L L L @{} >{\kern\tabcolsep} l @{}}    \toprule
\textbf{Parameters} & \textbf{Excitatory} &\textbf{Inhibitory} &\textbf{Units}  \\\midrule
$N_X$    & 12000  & 12000  & -    \\ 
\rowcolor{black!20}[0pt][0pt] $v_l^X$ & -70 & -70 & mV \\ 
\rowcolor{black!0}[0pt][0pt] $g_l^X$& 30 & 20 & nS \\
\rowcolor{black!20}[0pt][0pt] $C_X$ & 0.6 & 0.2 & nF \\ 
\rowcolor{black!0}[0pt][0pt] $v_t^X$& -50 & -50 & mV \\
\rowcolor{black!20}[0pt][0pt] $v_r^X$ & -90 & -60 & mV \\ 
\rowcolor{black!0}[0pt][0pt] $\mu_{bg}^X$& 320 & 50 & pA \\
\rowcolor{black!20}[0pt][0pt] $\sigma_{bg}^X$ & 9 & 9 & pA \\\bottomrule
 \hline
\end{tabular}

\caption[Integrate-and-Fire model parameters.] {\bf Integrate-and-Fire model parameters.}
 \label {tab:IFPar}
\end{table}

\begin{table}[H]%
\centering %
\begin{tabular}{@{} c L L L @{} >{\kern\tabcolsep} l @{}}    \toprule
\textbf{Parameters} & \textbf{E Presyn.} &\textbf{I Presyn.} &\textbf{Units}  \\\midrule
$\tau_r^{AMPA}$    & 0.5  & -  & ms    \\ 
\rowcolor{black!20}[0pt][0pt] $\tau_d^{AMPA}$ & 5 & - & ms \\ 
\rowcolor{black!0}[0pt][0pt] $g\tau_r^{NMDA}$& 1 & - & ms \\
\rowcolor{black!20}[0pt][0pt] $\tau_d^{NMDA}$ & 50 & - & ms \\ 
\rowcolor{black!0}[0pt][0pt] $\tau_r^{GABA_A}$& - & 1 & ms \\
\rowcolor{black!20}[0pt][0pt] $\tau_d^{GABA_A}$ & - & 5 & ms \\ 
\rowcolor{black!0}[0pt][0pt] $J_0^{YX}$& 20 & 18 & nC \\
\rowcolor{black!20}[0pt][0pt] $J_1^{YX}$ & 60 & 12.5 & nC \\\bottomrule
 \hline
\end{tabular}

\caption[Integrate-and-Fire model synaptic parameters.] {\bf Integrate-and-Fire model synaptic parameters.}
 \label {tab:IFParSyn}
\end{table}

\subsection*{Numerical methods}\label{sec:NumericalMethodsl}
Integration of the rate model equations was performed by the MATLAB ode45 solver, which uses explicit Runge-Kutta (4,5) formula with an adaptive time step. Bifurcation diagrams were computed using custom made scripts in MATLAB R2017b. Numerical solutions of equations were found with MATLAB fsolve, which implements a 'trust-region-dogleg' algorithm. Integration of the integrate-and-fire model was performed using the Euler method with a time step of 0.1 ms.

\newpage

\section*{Acknowledgments}
 We thank Carole Levenes and Carl van Vreeswijk for discussions.  This research was conducted within the scope of the France-Israel Laboratory of Neuroscience (CNRS) and supported by the Paris School of Neuroscience (ENP), UniNet EU excellence network,  the grants ANR-09-SYSC-002–01, ANR-14-NEUC-0001–01 , ANR-13-BSV4-0014-02 , and ANR-17-NEUC-0005. The funders had no role in study design, data collection and analysis, decision to publish, or preparation of the manuscript.

\section*{Author Contributions}
Conceived and designed the experiments: OH, AS,  DH. Performed the experiments: OH, AS, DH. Analyzed the data: OH, AS, DH. Contributed reagents/materials/analysis tools: OH, AS, DH. Wrote the paper: OH, AS, DH.

\bibliographystyle{unsrt}
\bibliography{MyBib}
\end{document}